\documentclass[useAMS,usenatbib]{mn2e}
\usepackage{graphicx}
\usepackage{color}
\usepackage{amssymb}
\usepackage{amsmath}
\usepackage{upgreek}
\usepackage{subfigure}
\usepackage{rotating}
\usepackage{pdflscape}
\usepackage{bm} 
\usepackage{mathtools}
\usepackage{epstopdf}
\usepackage{gensymb}

\usepackage{comment}
\excludecomment{details}
\graphicspath{{images/}}

\title[Galactic-scale hydrodynamic simulations of $^{26}$Al]{Synthetic $^{26}$Al emission from galactic-scale superbubble simulations}
\author[Rodgers-Lee, Krause, Dale, Diehl]{D. Rodgers-Lee$^{1,2}$\thanks{E-mail:
drodgers@tcd.ie}, M. G. H. Krause$^{1}$, J. Dale$^{1}$, R. Diehl$^{3}$ \\ 
$^{1}$ Centre for Astrophysics Research, School of Physics, Astronomy and Mathematics, University of Hertfordshire, College Lane,\\ Hatfield AL10 9AB, UK \\
$^{2}$ School of Physics, Trinity College Dublin, University of Dublin, College Green, Dublin 2, Co. Dublin, D02 PN40, Ireland \\
$^{3}$ Max-Planck-Institut f{\"u}r extraterrestrische Physik, Giessenbachstr. 1, 85741 Garching \\
}
\begin{document}
\date{Accepted xxxx xxxxxx xx. Received xxxx xxxxxx xx; in original form xxxx xxx xx}
\pagerange{\pageref{firstpage}--\pageref{lastpage}} \pubyear{xxxx}
\maketitle

\label{firstpage}

\begin{abstract}
Emission from the radioactive trace element $^{26}$Al has been observed throughout the Milky Way with the COMPTEL and INTEGRAL satellites. In particular the Doppler shifts measured with INTEGRAL connect $^{26}$Al with superbubbles, which may guide $^{26}$Al flows off spiral arms in the direction of Galactic rotation. In order to test this paradigm, we have performed galaxy-scale simulations of superbubbles with $^{26}$Al injection in a Milky Way-type galaxy.

We produce all-sky synthetic $\gamma-$ray emission maps of the simulated galaxies. We find that the 1809\,keV emission from the radioactive decay of $^{26}$Al is highly variable with time and the observer's position. This allows us to estimate an additional systematic variability of 0.2\,dex for a star formation rate derived from $^{26}$Al for different times and measurement locations in Milky Way-type galaxies. High-latitude morphological features indicate nearby emission with correspondingly high integrated $\gamma-$ray intensities. We demonstrate that the $^{26}$Al scale height from our simulated galaxies depends on the assumed halo gas density.

We present the first synthetic 1809\,keV longitude-velocity diagrams from 3D hydrodynamic 
simulations. The line-of-sight velocities for $^{26}$Al can be significantly different from the line-of-sight velocities associated with the cold gas. Over time, $^{26}$Al velocities consistent with the INTEGRAL observations, within uncertainties, appear at any given longitude, broadly supporting previous suggestions that $^{26}$Al injected into expanding superbubbles by massive stars may be responsible for the high velocities found in the INTEGRAL observations. We discuss the effect of systematically varying the location of the superbubbles relative to the spiral arms.

\end{abstract}

\begin{keywords}
hydrodynamics -- methods: numerical -- gamma-rays: ISM -- ISM: kinematics and dynamics -- galaxies: spiral -- stars: massive
\end{keywords}

\section{Introduction}
\label{sec:intro}

Massive star formation influences the evolution of disk galaxies via the collective power of stellar feedback in the form of superbubbles \citep{maclow_1988,de_avillez_2005,keller_2016,naab_2017}. The stellar energy output responsible for the formation of superbubbles is dominated by the contribution from massive stars, namely from stellar winds and supernovae. Superbubbles have been observed at many wavelengths: CO \citep{dawson_2013}, radio \citep{bagetakos_2011}, H$\alpha$ \citep{egorov_2017}, infrared \citep{ochsendorf_2015}, X-rays \citep{kavanagh_2012} and $\gamma$-rays \citep{hess_2015}. They may also be associated with the spiral arms of disk galaxies \citep{krause_2015}. Superbubbles may play an important role in the chemical enrichment of the intergalactic medium and are likely to be the driving force for galactic outflows \citep{heckman_2017}.

The radioisotope $^{26}$Al, with its decay lifetime of $10^{6}$yr, is an important tracer of massive star formation as it is thought to be produced by these stars and ejected into the interstellar medium (ISM) predominantly via stellar winds and supernovae \citep{prantzos_1996,diehl_2013}. It decays via $\beta-$decay producing a 1809\,keV $\gamma-$ray emission line. The radioactive decay time of $^{26}$Al \citep[$\tau_{1/2}\sim 7.17\times 10^{5}$\,yrs,][]{endt_1990} is comparable to the sound crossing time through the hot phase in superbubbles \citep{krause_2015} and therefore it traces the dynamics of superbubbles.

The COMPTEL observations have been transformed into a map \citep{pluschke_2001} which traces the 1809\,keV emission of $^{26}$Al in the Milky Way. This map shows emission that is centred on the Galactic plane with some identifiable features, such as the Cygnus star-forming region. Data from the SPI telescope \citep{vedrenne_2003} on the INTEGRAL satellite \citep{winkler_2003} have been used to confirm the association of $^{26}$Al with nearby massive star groups \citep{diehl_2010,martin_2010,siegert_2017,krause_2018}. \citet{kretschmer_2013} also investigated the kinematics of $^{26}$Al in the Galaxy using INTEGRAL data. They found that the observed velocities of the gas traced by $^{26}$Al were in excess of the velocities expected due solely to Galactic rotation. They were able to explain these observations by assuming that the $^{26}$Al sources were located along the inner spiral arms with a global blow-out preference in the forward direction. In an analytical model of the ISM, \citet{krause_2015} showed that massive stars are expected to form ahead of the gaseous spiral arms. The interaction of superbubbles, formed by the massive stars, with the gas then leads to $^{26}$Al outflows in the forward direction, where the observed velocity magnitude corresponds to the sound speed in the superbubbles.

There have been many simulations focusing on the influence of stellar feedback in galaxies. In recent years, these simulations have ranged from simulations focusing on individual superbubbles \citep{krause_2013,gupta_2018,gentry_2019} to 3D vertically stratified box simulations including stellar feedback in the form of supernovae \citep{de_avillez_2004,de_avillez_2005,girichidis_2016}. \citet{sarkar_2015} present 2D simulations similar to those presented in this paper focusing on galactic outflows. Only recently has it become possible to simulate entire galaxies including stellar feedback. \citet{fujimoto_2018} presented the first 3D galactic-scale hydrodynamic simulations of a spiral galaxy (similar in size to the Milky Way) which include self-gravity, stochastic star formation, H {\scshape ii} regions, supernovae and radioisotope injection. These simulations focus on the radioisotope abundances in comparison to the high radioisotope abundances found in the solar system relative to the ISM. They suggest that the seemingly high abundances are in fact typical and can be explained by star formation that is correlated on galactic scales.

In this paper we focus on the influence and kinematics of superbubbles on the surrounding ISM. We will investigate possible galactic outflows from our simulations in a separate paper. Section\,\ref{sec:form} presents our model for the galaxy and the initial conditions. We compare our simulations directly with COMPTEL and INTEGRAL $\gamma-$ray observations of the Milky Way in Section\,\ref{sec:results} and discuss our results in Section\,\ref{sec:discussion}. In Section\,\ref{sec:conclusions} we delineate our main findings and present our conclusions.

\section{Formulation}
\label{sec:form}

The simulations in this paper were run using the {\scshape pluto} code \citep{mignone_2007}. We model outflows from a galactic disk by solving the hydrodynamic equations on a Cartesian grid. The hydrodynamic equations solved are
\begin{flalign}
& \frac{\partial \rho}{\partial t} + \nabla \cdot (\rho\bm{v}) =\dot \rho_\mathrm{SB}\,;    \\
& \frac{\partial \rho \bm{v}}{\partial t} +\nabla \cdot (\rho\bm{v}\bm{v} + p\bold{I}) = -\rho\nabla \Phi  \,;\\
& \frac{\partial }{\partial t}(\rho e +\rho\Phi) + \nabla \cdot [(\rho e + p + \rho\nabla\Phi)\bm{v}e] = \dot e_\mathrm{SB}  
\end{flalign}
where $\rho, \bm{v}, p$ and $e$ are the mass density, velocity, pressure and internal energy density of the fluid, here corresponding to the gas in the galaxy. The gravitational potential, $\Phi$, is composed of a number of components that are described in the following section. The source terms $\dot \rho_\mathrm{SB}$ and $\dot e_\mathrm{SB}$ represent the injection of mass and internal energy associated with the superbubbles. The details of the superbubble implementation are given in Section\,\ref{subsec:sb-inj}.  

\subsection{Components of the gravitational potential}
\label{subsec:potentials}
The gravitational potential which we use as our model comprises a number of components: the disk ($\Phi_\mathrm{disk}$), central bulge ($\Phi_\mathrm{bulge}$), dark matter halo \citep[$\Phi_\mathrm{NFW}$, a Navarro-Frenk-White (NFW) potential,][]{navarro_1996} and spiral arms ($\Phi_\mathrm{spiral}$) such that 
\begin{equation}
\Phi = \Phi_\mathrm{disk}+\Phi_\mathrm{bulge}+\Phi_\mathrm{NFW}+\Phi_\mathrm{spiral}
\end{equation}

\subsubsection{Disk potential}
The disk potential we use is the same as the disk potential described in \citet{flynn_1996} which is a three component Miyamoto-Nagai potential \citep{miyamoto_1975}, given by
\begin{equation}
\Phi_\mathrm{disk} = \sum_{i=1}^3-\frac{GM_{\mathrm{D}_i}}{\sqrt{R^2 + \left (a_i +\sqrt{z^2+b^2}\right )^2}}  
\end{equation}
where $M_\mathrm{D_1}$, $M_\mathrm{D_2}$, $M_\mathrm{D_3}, a_1, a_2, a_3$ and $b$ are given in Table\,\ref{table:sim_parameters}. $R=\sqrt{x^2+y^2}$, where $x$ and $y$ are the usual Cartesian coordinates. 

\subsubsection{Bulge potential}
The central bulge potential is again from \citet{flynn_1996} and is described by
\begin{equation}
\Phi_\mathrm{bulge} = \sum_{i=1}^2-\frac{GM_{\mathrm{C}_i}}{\sqrt{r^2 +r_{\mathrm{C}_i}^2}}
\end{equation}
where $M_\mathrm{C_1},\, M_\mathrm{C_2},\,r_\mathrm{C_1}$ and $r_\mathrm{C_2}$ are given in Table\,\ref{table:sim_parameters} and $r=\sqrt{x^2+y^2+z^2}$, , where $z$ is the usual Cartesian coordinate.

\subsubsection{Dark matter halo potential}
\label{subsubsec:dark}
The dark matter halo potential, $\Phi_\mathrm{NFW}$, is given by a NFW profile,
\begin{equation}
\Phi_\mathrm{NFW} = -\frac{GM_{200}}{r_\mathrm{s}f(c_{200})}\frac{\mathrm{ln}(1+r/r_\mathrm{s})}{r/r_\mathrm{s}}
\end{equation}
where the function $f(c_{200})=\mathrm{ln}(1+c_{200})-c_{200}/(1+c_{200})$, $M_{200}=1\times 10^{12}M_\odot$ \citep{taylor_2016} and $c_{200}$ is the concentration parameter. The concentration parameter is defined as $c_{200}=r_{200}/r_\mathrm{s}$ where $r_{200}=258\,$kpc and $r_\mathrm{s}$ is the scale radius. The scale radius has an influence on the structure of the hydrostatic gas halo of a galaxy \citep[compare with][]{krause_2019}. 

For our main setup the physical parameters we choose are based on X-ray observations of \citet{gupta_2017} which probe the circumgalactic medium of the Galaxy. They find a large gas scale height, $>40\,$kpc, and a low value for the halo density (discussed in Section\,\ref{subsubsec:rho}). We therefore adopt a scale radius of $r_\mathrm{s}=40$\,kpc which implies a concentration factor $c_{200}=6.45$, and is compatible with the observational values of \citet{klypin_2016} for galaxies with masses similar to the Milky Way. 

In Section\,\ref{sec:discussion} we discuss simulations which use a different choice of $r_\mathrm{s}$ since there are significant uncertainties regarding the structure of the Milky Way's hot-gas halo \citep{gupta_2017,bregman_2018}.

\subsubsection{Spiral arm potential}
\label{subsubsec:sap}
The spiral arm potential we include in our model is taken from \citet{dobbs_2006}, based on the potential from \citet{cox_2002}, and given by
\begin{eqnarray}
\Phi_\mathrm{sp}(R,\theta,t) = -4\pi GH_\mathrm{s}\rho_{\mathrm{sp},0}\,\mathrm{exp}\left(-\frac{R-R_0}{R_\mathrm{s}} \right) \nonumber \\
\times \sum\limits_{n=1}^3 \frac{C_n}{K_nD_n}\mathrm{cos}(n\gamma)
\end{eqnarray}
where $H_s=0.18\,$kpc is the scale height of the disk, $\rho_{\mathrm{sp},0}=14m_\mathrm{H}/11$, $C_1=8/(3\pi), C_2=1/2$ and $C_3=8/(15\pi)$. The values of $R_0$ and $R_s$ are given in Table\,\ref{table:sim_parameters}. The two functional parameters are given by
\begin{eqnarray}
K_n &=& \frac{nN}{R\mathrm{sin}(\alpha)}, \\
D_n &=& \frac{1+K_nH+0.3(K_nH)^2}{1+0.3K_nH}
\end{eqnarray}
\noindent where $N=4$ is the number of spiral arms and $\alpha=15\degree$ is the pitch angle of the spirals, similar to the Milky Way \citep{vallee_2005}. Finally,
\begin{equation}
\gamma = N\bigg[\theta - \Omega_\mathrm{p}t -\frac{\mathrm{ln}(R/R_0)}{\mathrm{tan(\alpha)}}\bigg] 
\end{equation}
\noindent where $\Omega_\mathrm{p}=2.0\times10^{-8}\,\mathrm{rad \,yr^{-1}}$ is the pattern speed.

\begin{table}
\centering
\caption{List of parameters for simulations}
\begin{tabular}{@{}llr@{}}
\hline
Component & parameter & value \\
\hline
$\Phi_\mathrm{disk}$ & $b$ & 0.3\,kpc \\
& $a_1$ & 5.81\,kpc \\
& $M_\mathrm{D_1}$ & $6.6\times 10^{10}M_\odot$ \\
& $a_2$ & 17.43\,kpc \\
& $M_\mathrm{D_2}$ & $-2.9\times 10^{10}M_\odot$ \\
& $a_3$ & 34.86\,kpc \\
& $M_\mathrm{D_3}$ & $3.3\times 10^9M_\odot$ \\

$\Phi_\mathrm{bulge}$ & $r_\mathrm{C_1}$ & 2.7\,kpc \\
& $M_\mathrm{C_1}$ & 3.0$\times 10^{9}M_\odot$ \\
& $r_\mathrm{C_2}$ & 0.42\,kpc \\
& $M_\mathrm{C_2}$ & $1.6\times10^{10}M_\odot$ \\
$\Phi_\mathrm{NFW}$ & $r_\mathrm{s}$ (lower $\rho_\mathrm{halo,0}$) & 40.0\,kpc \\
										& $r_\mathrm{s}$ (higher $\rho_\mathrm{halo,0}$) & 21.5\,kpc \\
$\Phi_\mathrm{spiral}$ & $R_0$	& 8.0\,kpc \\
										& $R_\mathrm{s}$ & 7.0\,kpc \\
\hline
\end{tabular}
\label{table:sim_parameters}
\end{table}

\subsection{Initial conditions for model of Milky Way-type galaxy}
\label{subsec:initial_conditions}
The external boundary conditions of the computational domain are set to be outflow boundary conditions. A uniform and stretched mesh Cartesian grid is used for the simulations to maximise the resolution in the galactic disk. In the stretched grid the cell size increases outwards by a constant ratio between consecutive cells. The overall extent of the simulations is -100 to 100 kpc in all directions. The stretched grid ranges from -100 to -11\,kpc and 11-100\,kpc in the $x$ and $y$ directions with 30 grid zones for each side. The uniform grid in these directions therefore ranges from -11 to 11\,kpc with a spatial resolution of 0.125\,kpc (the number of cells is $N_{x,y}=176$). In the $z$ direction the stretched grid ranges from -100 to -0.25\,kpc and 0.25 to 100\,kpc with 37 grid zones for each side. The uniform grid ranges from -0.25 to 0.25\,kpc with a spatial resolution of $\sim$0.02\,kpc (the number of cells is $N_{z}=23$).  

\subsubsection{Temperature and density profile}
\label{subsubsec:rho}
The equation of state for an ideal gas is used for the simulations, as appropriate for modelling the majority of the gas in the ISM, namely atomic hydrogen. The initial temperature in the disk is chosen to be $T_\mathrm{disk}=10^4\,$K and the initial temperature in the halo is taken to be $T_\mathrm{halo}=2\times 10^6\,$K, based on the recent X-ray observations of \citet{gupta_2017}. 

\begin{figure}
\centering
 \includegraphics[width=0.5\textwidth]{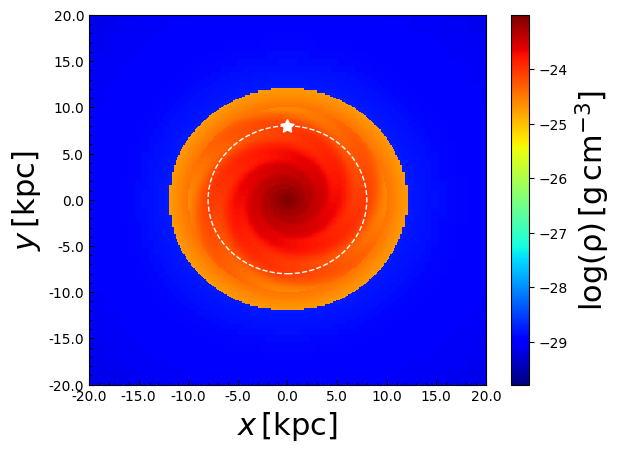}
  \caption{The $xy$-plane (for $z=0$) of the initial gas density of the galaxy, given in Eq.\,\ref{eq:rho}. The dotted white circle indicates the approximate galactocentric radius of the solar system which is one of the assumed locations of the observer used for the synthetic $\gamma-$ray emission maps shown in Section\,\ref{sec:results}. }
\label{fig:rho-initial}
\end{figure}

The initial mass density profile of the system is set up in a way similar to \citet{von_glasow_2013}, except for the fact that we use physical parameters relevant for a Milky Way-type galaxy rather than a Lyman-break galaxy. The initial mass density profile for the galaxy is shown in Fig.\,\ref{fig:rho-initial}. As in \citet{von_glasow_2013} there are contributions from the halo ($\rho_\mathrm{halo}$) and the disk ($\rho_\mathrm{disk}$) but now there is an additional contribution from the spiral arm density profile ($\rho_\mathrm{spiral}$) such that
\begin{equation}
\rho = \rho_\mathrm{disk} + \rho_\mathrm{halo} +\rho_\mathrm{spiral}
\label{eq:rho}
\end{equation}
where the disk density is given by
\begin{equation}
\rho_\mathrm{disk} = 
\begin{cases}
\rho_\mathrm{disk,0}\,\mathrm{exp}\left (-\frac{R}{r_\mathrm{s,D}}\right ) & \text{if } R<12\text{kpc} \text{ and } |z|<500\text{pc}\\
\hspace{10mm}0			& \text{if } R>12\text{kpc} \text{ or } |z|>500\text{pc} \\
\end{cases}
\end{equation}
where $\rho_\mathrm{disk,0}=1\times 10^{-23}\mathrm{g\,cm^{-3}}$ and the disk scale radius, $r_\mathrm{s,D}=3.0$\,kpc. This gives an initial gaseous disk mass of $M_d = 1.1\times 10^{10}M_\odot$, similar to the total gas mass in the Milky Way as obtained from HI observations \citep{kalberla_2008,nakanishi_2016}.

\noindent The halo density profile is set up in hydrostatic equilibrium and is defined as 
\begin{equation}
\rho_\mathrm{halo} = \rho_\mathrm{halo,0}\,\mathrm{exp}\left( -\Phi\frac{0.59m_\mathrm{p}}{k_\mathrm{B}T_\mathrm{halo}}\right)
\label{eq:rho-halo}
\end{equation}
where $m_\mathrm{p}$ is the proton mass and $k_\mathrm{B}$ is the Boltzmann constant. For the main setup we adopt $\rho_\mathrm{halo,0} =4.0\times 10^{-31}\mathrm{g\,cm^{-3}}$ which is compatible with $\rho_\mathrm{halo}(r=0) = 4.4\times 10^{-28}\,\mathrm{g\,cm^{-3}}$ \citep[][their central value]{gupta_2017}. 

\noindent The initial spiral arm density profile is similar to that presented in \citet{cox_2002}, without the vertical component, given by
\begin{equation}
\rho_\mathrm{spiral}(R,\theta) = \rho_{\mathrm{sp},0}\,\mathrm{exp\left(-\frac{R-R_0}{R_\mathrm{s}} \right)}\mathrm{cos(\gamma)}
\end{equation}
\noindent where the definition of $\gamma$ and values of $R_0$ and $R_\mathrm{s}$ are given in Section\,\ref{subsubsec:sap} and Table\,\ref{table:sim_parameters}. Here $\rho_{\mathrm{sp},0}=m_\mathrm{H}/11$ is used.

\subsubsection{Velocity profile}
The velocity field of the galactic disk is set up initially as Keplerian with
\begin{equation}
v_\mathrm{\phi} = \sqrt{R\frac{d\Phi}{dR}}
\end{equation}
\noindent The halo material is set to initially have zero velocity.

\subsection{Radiative cooling}
We include time-dependent optically thin radiative losses using {\scshape pluto}'s tabulated cooling function to update the internal energy such that
\begin{equation}
\frac{\partial \rho e}{\partial t} = -n^2\Lambda(T)
\end{equation}
where $n=\frac{\rho}{\mu m_u}$ is the number density and $\Lambda(T)$ is the cooling function. The cooling function is constrained to only operate above $1.2\times 10^{4}\,$K. As described by \citet{sarkar_2015} this approximation indirectly accounts for the continuous stellar heating of the gas in the disk that we do not include. The cut-off in the cooling function prevents the formation of cold gas clouds which would be unresolved at least in parts of our simulations.

\subsection{Superbubble implementation}
\label{subsec:sb-inj}
In this paper our main aim is to study the influence of superbubbles on a Milky Way-type disk galaxy. Therefore, rather than modelling each massive star separately we use a sub-grid model (which we describe in some detail below) to simulate the influence of a cluster of massive stars which have formed a superbubble. This allows us to simulate a full 3D disk galaxy computationally. 

Star formation in the Milky Way is clustered in a hierarchical way \citep[e.g.,][]{krumholz_2014}. Similarly, bubbles merge continuously into bigger bubbles \citep[e.g.,][]{krause_2015, krause_2018}.  We include superbubbles corresponding to a stellar content of $10^6 M_\odot$ and any further merging is then followed by the simulation.
\subsubsection{Energy and mass injection rates of superbubbles}
For each superbubble in our simulations we assume the young cluster would contain $\sim 10^6$ stars. Therefore, considering a Salpeter-type initial mass function \citep{salpeter_1955}, $\sim 10^{4}$ of these stars will be massive stars. Each massive star injects $10^{36}\mathrm{erg\,s^{-1}}$ \citep[as summarised from the stellar evolution models of][]{voss_2009}, therefore $\dot e_\mathrm{SB} = 10^{40}\mathrm{erg\,s^{-1}}/V_\mathrm{inj}$ is injected per superbubble comprising $10^{4}$ massive stars, where $V_\mathrm{inj}$ is the volume of the injection region. For each superbubble a corresponding mass density injection rate of $\dot \rho_\mathrm{SB} = 10^{24}\mathrm{g\,s^{-1}}/V_\mathrm{inj}$ is also injected \citep[following][]{sarkar_2015}. The injection volume, $V_\mathrm{inj}$ is defined as a sphere of radius, $r_\mathrm{inj}$. \citet{krause_2014} fitted an analytic superbubble model to 3D hydrodynamic simulations with detailed injection histories of up to three massive stars and found 
$r_\mathrm{inj} \sim 500\,\mathrm{pc}\,(L/10^{40}\mathrm{erg\,s^{-1}})^{1/5} (\rho_0/1.67\times 10^{-24}\mathrm{g\,cm^{-3}})^{-1/5} (t/10\,\mathrm{Myr})^{3/5}$ for the superbubble radius. We chose $r_\mathrm{inj}=300$\,pc to ensure sufficient resolution of the injected bubbles, as well as to reflect the expected size of the superbubble corresponding to a star forming region of the chosen size.

We inject the cumulative energy and mass input from the first 10\,Myr of the stellar population lifetime \citep[see][for a population synthesis treatment of massive star groups]{voss_2009} instantaneously at the initial injection time for each superbubble and then the superbubble continues to inject mass and energy at the aforementioned rates for another 25\,Myr, equivalent to the typical lifetime of $\sim 35$\,Myr. This reflects the fact that we do not attempt to capture the early merging of bubbles in a star forming region, but rather pick up the evolution after a certain amount of time.

\subsubsection{Temporal and spatial injection of superbubbles}
The superbubbles are injected randomly in time. Specifically, 833 random initial injection times are selected between 0-250\,Myr. This corresponds to a typical star formation rate of 3 $M_\odot\mathrm{yr^{-1}}$ in comparison to the Milky Way's star formation rate which is estimated to be $\sim 1-4 M_\odot\mathrm{yr^{-1}}$ \citep{diehl_2006,chomiuk_2011,licquia_2015}. We chose the higher value to be better able to address derivations of the star formation rate from $^{26}$Al: \citet{diehl_2006} had estimated a star formation rate of $4\,M_\odot \mathrm{yr}^{-1}$, which was recently updated to $3\,M_\odot \mathrm{yr}^{-1}$ by taking better account of local foreground. We discuss this in more detail below. We take 35\,Myr to be the lifetime of the superbubbles which corresponds to the lifetime of $9 M_\odot$ stars \citep{georgy_2013}. 

For the Milky Way the star formation rate surface density increases towards the Galactic centre \citep[see Fig.7 of][for instance]{kennicutt_2012}. The Milky Way also has a negative metallicity gradient as a function of galactocentric radius and $^{26}$Al yields from the winds of massive stars should increase with stellar metallicity. For the simulation that we will refer to as `the fiducial run' throughout the paper, the superbubbles are injected along the spiral arms of the galaxy, specifically at the position of maximum spiral arm density (of the initial density profile rotated as appropriate for the current simulation time) at any given radius. All the superbubbles are initialised at the mid-plane of the disk at $z=0$. The galactocentric injection radius, $R_\mathrm{SB}$, of the superbubble is chosen as a random number between $0-10\,$kpc. This implies, due to geometric dilution, that the star formation rate surface density in the simulations decreases as a function of radius and is proportional to 1/R. Given a specific value of $R_\mathrm{SB}$, the angle, $\theta_\mathrm{SB}$, is then selected such that the superbubble lies on a spiral arm. 

It is important to note that superbubbles are thought to form from Wolf-Rayet winds which would be expected to become effective $\sim$3\,Myr after  the stars' formation which took place on the spiral arms of galaxies. Thus, superbubbles should be offset from the maximum spiral arm density in the forward direction \citep{krause_2015}. Therefore, we run two other simulations where the superbubbles are injected 0.1 radians ahead of, and behind (as a reference for comparison), the position of the maximum spiral density, respectively. This allows us to investigate the dependence of the kinematic signature of the superbubbles on their injection position. We chose 0.1 radians based on the idea that young star clusters form on the spiral arms of galaxies. Considering the pattern speed of the Milky Way given above as $\Omega_\mathrm{p}=2.0\times10^{-8}\mathrm{rad \,yr^{-1}}$ implies that in $\sim 5$\,Myr the stars would drift 0.1 radians relative to the spiral arms as they begin to form a superbubble.

Note that the initial galactic disk radius is 12\,kpc, slightly larger than the maximum injection radius of the superbubbles. We found that if the superbubbles were injected too close to the edge of the Galactic disk that, due to the low density surroundings, they were able to quickly expand into the non-rotating halo. This led to an exchange of angular momentum with the halo leading to a rapid shrinking of the galactic disk. A similar process is described in \citet{elmegreen_2014}. Star formation would be more naturally regulated in the outer regions of the disk than in our simulations.

\subsection{$^{26}$Al tracer fluid}
\label{subsection:isotope}
The radioactive isotope $^{26}$Al is ejected in the Wolf-Rayet phase of massive star evolution and by supernovae \citep{prantzos_1996,diehl_2013}. It decays predominantly via $\beta$ decay producing a 1809\,keV $\gamma$-ray photon, providing a tracer of massive star formation. The COMPTEL 1809\,keV survey \citep{pluschke_2001} maps this emission. By including a tracer fluid in our simulations to represent $^{26}$Al we can compare our results with the COMPTEL observations. This tracer fluid is a passive scalar which is advected along with the fluid representing the gas in the galaxy. $^{26}$Al decays with a half-life of $\sim 7.17\times 10^{5}$\,yr which we account for in order to evaluate the mass of $^{26}$Al in the tracer fluid. 

\noindent The tracer fluid is injected at the same time as each superbubble is injected. The mass injection rate as a function of time per superbubble is given by
\begin{equation}
\dot M(t) = 10^4\frac{t^2}{20+t^3} M_\odot\, \mathrm{Myr^{-1}}
\label{eq:m26}
\end{equation}
\noindent which was chosen, after including the effect of radioactivity, to match the massive star population synthesis results from \citet{voss_2009}. It is important to note that the mass injection rate in Eq.\,\ref{eq:m26} corresponds to the rate expected for the entire superbubble containing $10^4$ massive stars rather than the average per star as in \citet{voss_2009}. The mass injection rate in Eq.\,\ref{eq:m26} per star (obtained by simply dividing by $10^4$) convolved with radioactive decay as a function of time is shown in Fig.\,\ref{fig:isotope}.

\section{Results}
\label{sec:results}

Here we present the results from our 3D hydrodynamic simulations. We examine 3 simulations which vary the injection position of the superbubbles relative to the spiral arms of the galaxy. Throughout the paper, the simulation with the superbubble injection centred on the spiral arms is referred to as the fiducial simulation. In Section\,\,\ref{sec:discussion} we discuss the effect of assuming a higher halo density for the galaxies. 

\subsection{Time evolution of the superbubbles}
\begin{figure*}
\centering
 \includegraphics[width=\textwidth]{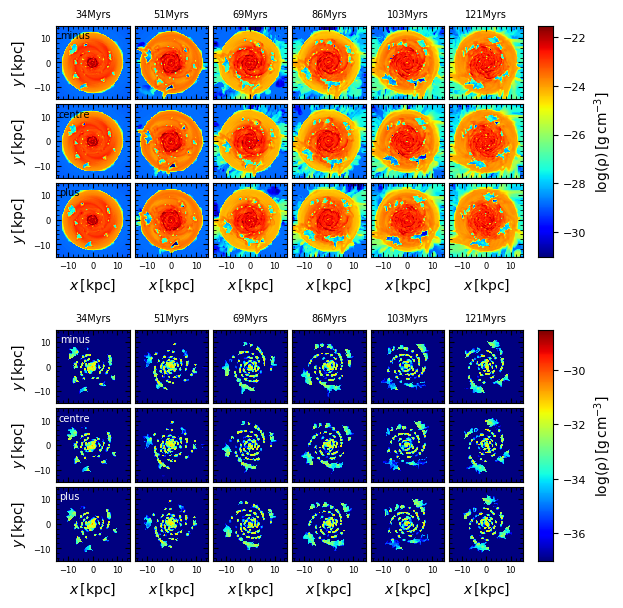}
  \caption{Mass density plots for the 3 simulations (with $\rho_\mathrm{halo,0}=4.0\times 10^{-31}\mathrm{g\,cm^{-3}}$) as a function of time. The top three panels show the mid-plane values of the $xy$-plane for the gas mass density and the bottom three panels show the corresponding $^{26}$Al mass density. In both the top and bottom panel the simulation data with the superbubbles injected behind, on and ahead of the spiral arms are shown in the top, middle and bottom row of images, respectively. As mentioned in Section\,\ref{subsec:initial_conditions} the extent of the grid is 100\,kpc in all directions. These plots show only a part of the computational domain focusing on the galactic disk.}
\label{fig:rho-panel}
\end{figure*}

\begin{figure*}
\centering
 \includegraphics[width=\textwidth]{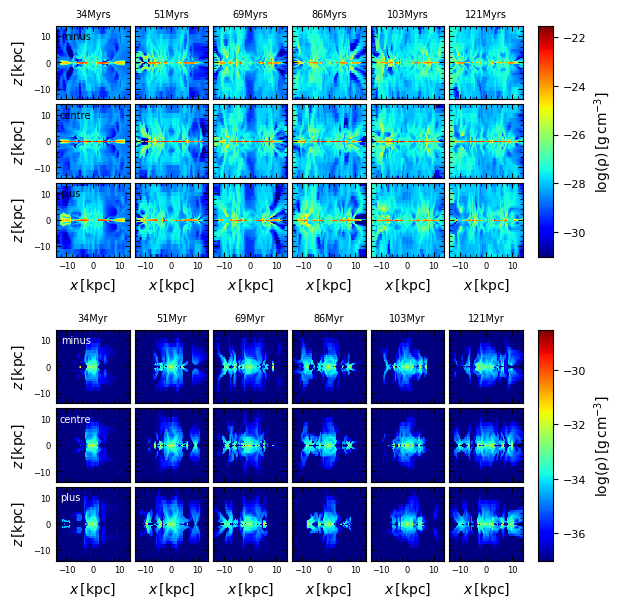}
  \caption{Mass density plots of the 3 simulations (with $\rho_\mathrm{halo,0}=4.0\times 10^{-31}\mathrm{g\,cm^{-3}}$) as a function of time. Same as for Fig.\,\ref{fig:rho-panel} except now showing the $xz-$plane values for $y=0$. }
\label{fig:rho-panel-slice}
\end{figure*}

In order to examine the time evolution of the galactic disk and superbubbles we plot the gas and $^{26}$Al mass densities for each of the three simulations as a function of time in Figs.\,\ref{fig:rho-panel}-\ref{fig:rho-panel-slice} between $t=34-121\,$Myr. The top three panels in Fig.\,\ref{fig:rho-panel} show the gas density in the $xy$-plane for $z=0$ while the bottom three panels show the corresponding $^{26}$Al densities. Fig.\,\ref{fig:rho-panel-slice} shows the same quantities in the $xz$-plane for $y=0$.

\begin{figure}
\centering
 \includegraphics[width=0.5\textwidth]{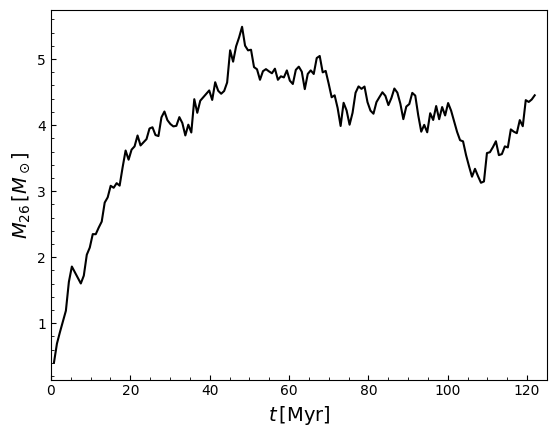}
  \caption{Total mass of $^{26}$Al in the galaxy for the fiducial simulation as a function of time.}
\label{fig:m26}
\end{figure}

By $t=34\,$Myr the superbubbles have begun to disrupt the galactic disk. There are no significant differences seen between the 3 simulations. The vertical slices, shown in Fig.\,\ref{fig:rho-panel-slice}, suggest that the simulations are well in dynamical equilibrium by 69\,Myr. The same impression is reached if one inspects the total $^{26}$Al mass over time (Fig.~\ref{fig:m26}). Therefore, all time-averaging of quantities and statistical analysis in the following sections is performed between $t=69$\,Myr until the end of the simulations at $t=121\,$Myr. In all three simulations the superbubbles disrupt the galactic disk and cause large low density regions above and below the disk (see Fig.\,\ref{fig:rho-panel-slice}).  


\subsection{Synthetic $\gamma-$ray maps}
\label{subsec:maps}

\subsubsection{Calculation of $\gamma-$ray emission}
In order to compare our simulation results with the COMPTEL map, which results from observations with an instrument that has an angular resolution of 3.8$^\circ$ \citep{pluschke_2001}, we create synthetic $\gamma-$ray maps of the simulated galaxies by integrating our result cubes along the lines of sight using the Hammer map projection. We chose the position of the observer to be located at a galactocentric radius of 8\,kpc which is approximately the same galactocentric radius as the Earth is within our Galaxy \citep[8.32\,kpc,][]{gillessen_2017}. 

\noindent The $\gamma-$ray flux density, $F_\gamma(x,y,z)$, from each emitting cell on the computational grid seen by an observer is 
\begin{equation}
F_\gamma = \frac{1809\mathrm{keV}}{4\pi d^2}\frac{dn_\gamma}{dt}dV
\label{eq:flux_density}
\end{equation}
where $d(x,y,z)$ is the distance from the observer to the source of emission, $n_{\gamma}(x,y,z)$ is the number density of $\gamma$-rays and $dV(x,y,z)$ is the volume of the emitting region. Considering a time, $t$, where the number density of $^{26}$Al is $n_{26}$, then $n_{\gamma}$ due to radioactive decay can be calculated as
\begin{equation}
\frac{dn_\gamma}{dt} = -\frac{dn_{26}}{dt} =  \frac{n_{26}}{\tau}
\end{equation}
where the mean half-life, $\tau=1.034$\,Myr. To produce a 2D emission map from the 3D set of fluxes given by Eq.\,\ref{eq:flux_density} we bin the data as a function of galactic longitude and latitude using the same resolution as for COMPTEL map ($\sim4^\circ$) and sum up the fluxes in each bin.  

It is important to note that our galactic-scale simulations are not able to include specifics such as the relative location of the Sun with respect to the spiral arms and nearby stellar groups. Therefore, it is not possible to make a direct comparison between our simulated galaxy and the COMPTEL map of the Milky Way. Instead, by examining different viewing angles and snapshots in time, we can identify synthetic emission maps that look similar to the COMPTEL map and examine the properties of those maps.

It is also worth noting that aliasing is introduced as we convert from a Cartesian grid to the emission maps. In particular, emission close to the observer contributes to a scalloping pattern as noted in \citet{fujimoto_2018}. Although we could exclude nearby emission from these maps, either simply by proximity to the observer or by a certain per cent of the emission closest to the observer, we present the maps unaltered instead. We have also included two examples in Appendix\,\ref{appendix:maps} which illustrate the effect of the aliasing in specific cases. The main point to note is, from Fig.\,\ref{fig:full_emission}, that the aliasing will create artificial low and high emission regions at intervals of 45$^\circ$ in longitude, centred on the galactic centre at 0$^\circ$. More details are given in Appendix\,\ref{appendix:maps}.

In the following sections we will describe the variation we observe due to viewing angle (Fig.\,\ref{fig:gamma_different_angles}) and the implications this has for any star formation rate derived from $\gamma-$ray emission maps. We present a panel of maps with the emission binned for different distances from the observer which shows the distance at which the majority of the emission originates (Figs.\,\ref{fig:panel_spatial}-\ref{fig:cum-both}). We also examine the time variability of the emission in Figs.\,\ref{fig:median}-\ref{fig:histogram}.

\subsubsection{Spatial variation of $\gamma-$ray emission as a function of viewing angle}
\label{subsubsec:spatial-variation}
\begin{figure*}
\centering
 \includegraphics[width=\textwidth]{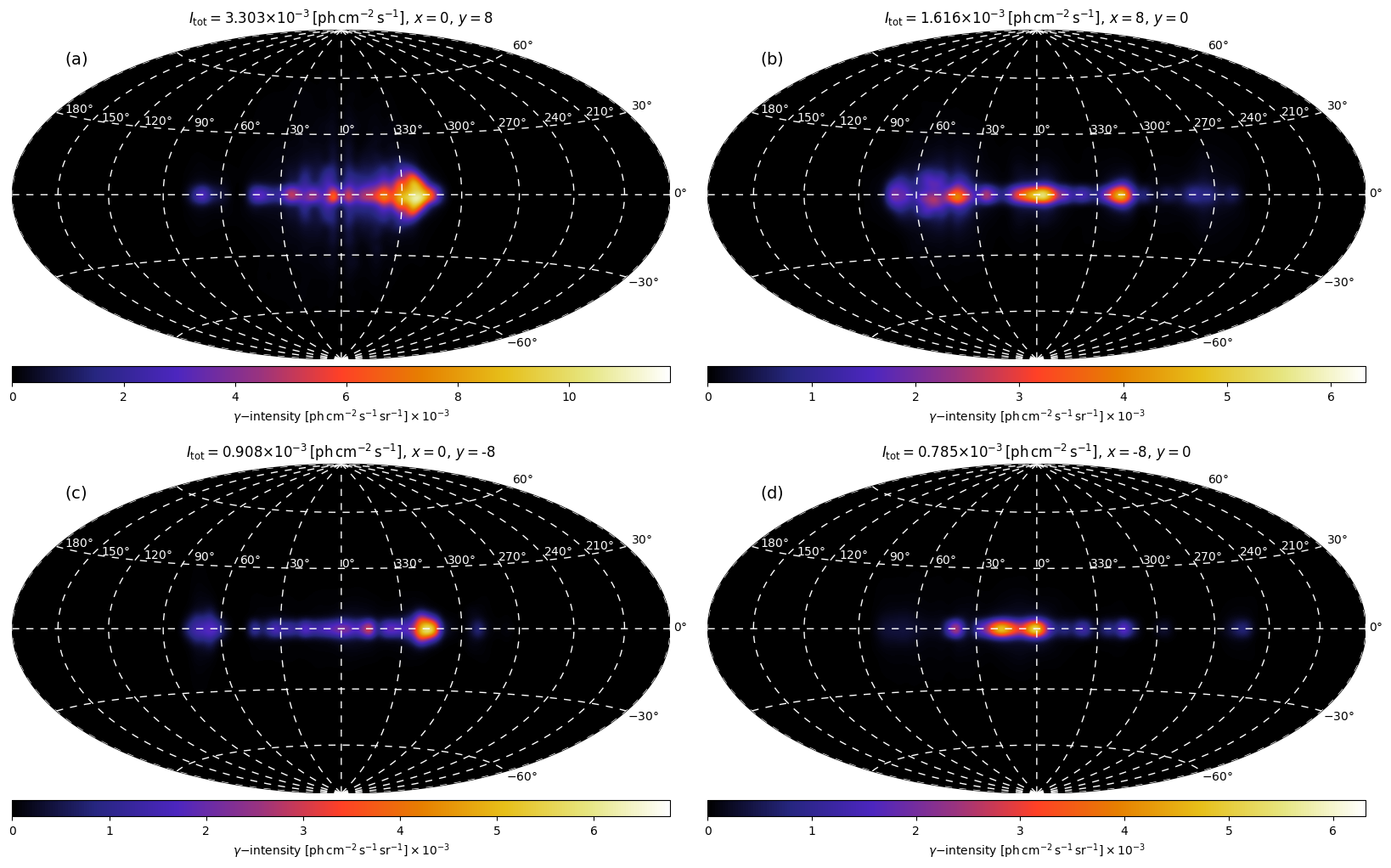}
  \caption{Synthetic maps of the 1809\,keV emission for the galaxy from the fiducial simulation at $t=103.8$\,Myr (shown in the fifth column and second row of the bottom panel in Fig.\,\ref{fig:rho-panel}) for different viewing angles located at the same distance from the Galactic centre. The total integrated flux density is given above each image, as well as the Cartesian coordinates of the observer (in kpc).}
\label{fig:gamma_different_angles}
\end{figure*}

We present a view of the fiducial simulation of the galaxy from 4 different angles in Fig.\,\ref{fig:gamma_different_angles} at $t=103.8\,$Myr. The viewing angles are every $90^\circ$ degrees. The Cartesian coordinates and the total $\gamma-$ray intensity for each viewing angle are given above each map.

The morphology of the emission differs as a function of viewing angle. We can compare our synthetic observations with the COMPTEL map, specifically the lower panel of Fig.\,5 from \citet{pluschke_2001}. Particular features, such as the Cygnus star-forming region located at 90$^\circ$ longitude, have been identified in the COMPTEL map. Fig.\,\ref{fig:gamma_different_angles}\,(a) looks most morphologically similar to the COMPTEL map, except the large extended emission is located at 330$^\circ$ longitude rather than 90$^\circ$. The fact that the large extended emission is not a universal feature for different viewing angles suggests that this emission is close-by in origin which we discuss in the following Section\,\ref{subsubsec:spatial-bins}. Our view of the Galaxy is naturally unique and dominated by the contribution from local sources.

\subsubsection{Spatial binning}
\label{subsubsec:spatial-bins}
\begin{figure*}
\centering
 \includegraphics[width=\textwidth]{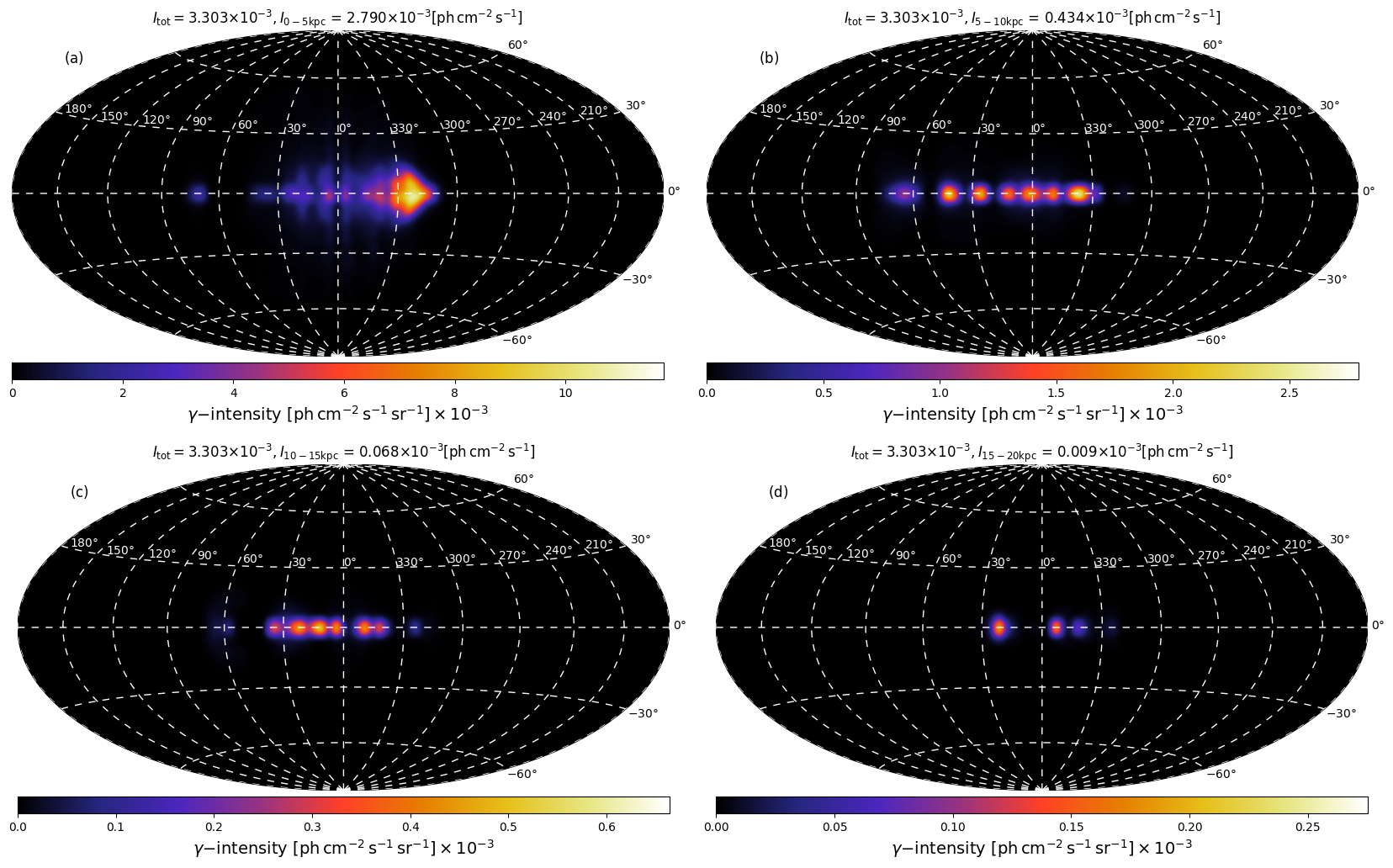}
  \caption{ $\gamma-$ray emission maps for the fiducial simulation at $t=103.8\,$Myr using different distance bins. The distance bins are 0-5, 5-10, 10-15 and 15-20\,kpc corresponding to a,\,b,\,c,\,d respectively. Above each plot the total integrated flux density is given, as well as the integrated flux density value for the distance bin used. }
\label{fig:panel_spatial}
\end{figure*}

Here, we bin the $\gamma-$ray emission as a function of distance from the observer, using the same viewing angle as presented in Fig.\,\ref{fig:gamma_different_angles}\,(a). The distance bins we use are 0-5, 5-10, 10-15 and 15-20\,kpc corresponding to Fig.\,\ref{fig:panel_spatial}\,(a)-(d), respectively. These images clearly show that the most spatially extended emission originates relatively close to the observer (within 5\,kpc). The emission from larger distances is also located closer to the Galactic centre with less emission coming from larger longitudes or latitudes. The large scale emission shown in Fig.\,\ref{fig:panel_spatial}\,(a) can be associated with the superbubble visible in Fig.\,\ref{fig:rho-panel} (top panel, middle row, fifth column) which lies within $\sim$1\,kpc of the observer at $x=0$ and $y=8$\,kpc, similar to the Cygnus star-forming region in the COMPTEL map.

\begin{figure*}%
	\centering
    \subfigure[ ]{%
        \includegraphics[width=0.5\textwidth]{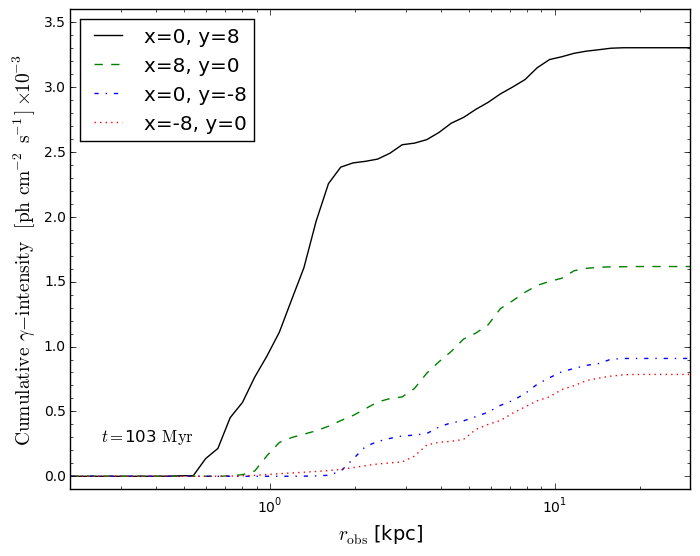}
       	\centering
\label{fig:cum}}%
  	    ~
    \subfigure[ ]{%
        \includegraphics[width=0.5\textwidth]{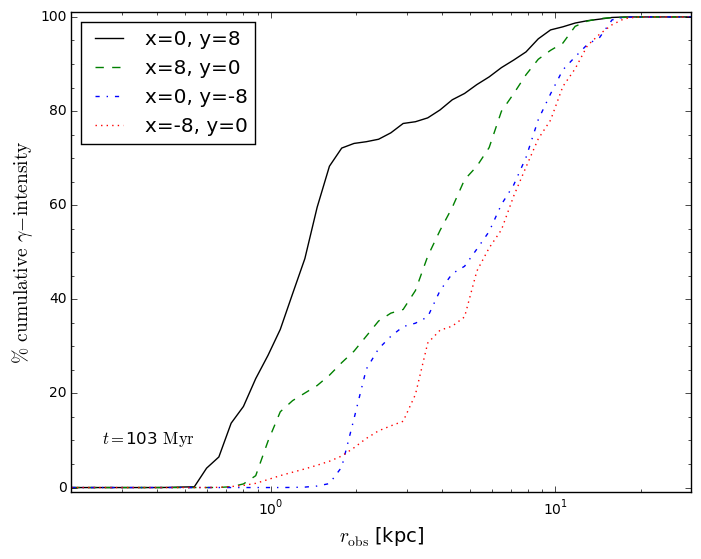}
	\centering
 \label{fig:cum_percent}}%
    \caption{(a) plots the cumulative flux and (b) plots the cumulative flux as a per cent of the total integrated flux as a function of distance from the observer for four different viewing angles (as indicated on the plots) in the galaxy at $t=103$Myr  for the fiducial simulation. } 
    \label{fig:cum-both}%
\end{figure*}

To investigate the variation with viewing angle more qualitatively, we examine the cumulative flux density as a function of distance from the observer for the four different viewing angles shown in Fig.\,\ref{fig:cum}. Fig.\,\ref{fig:cum_percent} similarly plots the cumulative flux density as a per cent of the integrated flux as a function of distance.

The different rates of increase in the cumulative flux density as a function of distance between the viewing angles indicate that large differences in the total $\gamma-$ray intensity will be due to local sources.
Fig.\,\ref{fig:cum} shows that, irrespective of the viewing angle, from $\gtrsim 1.5$\,kpc the addition to the cumulative flux is comparable.

The solid black line in Fig.\,\ref{fig:cum_percent} illustrates that $\sim 80\%$ of the total $\gamma-$ray intensity originates within 1.5\,kpc of the observer for this viewing angle (corresponding to Fig.\,\ref{fig:gamma_different_angles}\,(a)). This particular viewing angle shows corresponding large scale emission in the synthetic emission maps, as noted above. This would suggest that high values for the total intensity and emission at large latitudes are correlated with nearby emission. This plots also indicates that, irrespective of viewing angle, nearly all of the observed emission originates within $\sim 10$\,kpc of the observer.

\subsubsection{Time evolution}
\begin{figure}
\centering
 \includegraphics[width=0.5\textwidth]{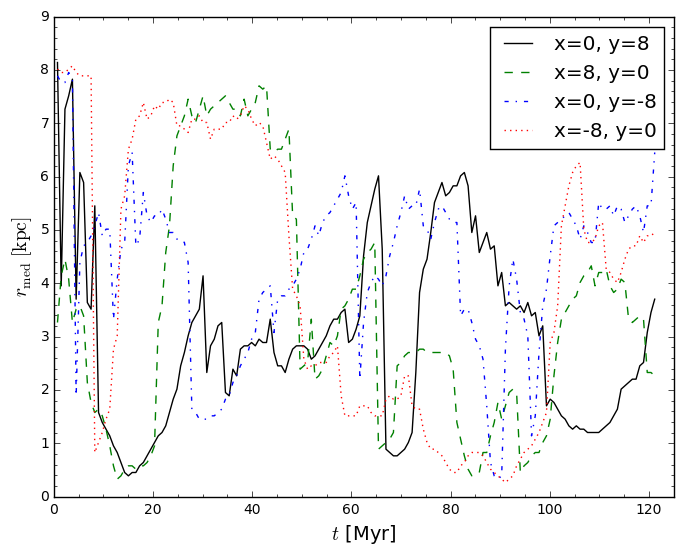}
  \caption{Median flux density distance for the fiducial simulation plotted as a function of time for four different viewing angles.}
\label{fig:median}
\end{figure}

The synthetic $\gamma-$ray emission maps are also variable with time, similar to the spatial variation discussed in Section\,\ref{subsubsec:spatial-variation}. We present a series of emission maps in Fig.\,\ref{fig:panel_time} to illustrate this. To quantify the time variability further we examined the median flux density distance ($r_\mathrm{med}$, the distance within which 50\% of the received flux originates) over time for the four different viewing angles considered, shown in Fig.\,\ref{fig:median}.  It is important to note that due to the finite grid size it is not possible to calculate the distance where exactly 50\% of the emission originates. Instead we plot the closest percentage allowed by the grid without interpolating. In general, the values are within 2\% of the actual median with some outliers. Fig.\,\ref{fig:median} clearly shows that, irrespective of viewing angle, the median flux density distance can easily vary by a factor of five on time scales of a few Myr.

\subsubsection{Integrated flux density values}
\label{subsubsec:int-flux}
The integrated flux density for the inner region of the Milky Way ($|\ell| \leq 30^\circ, \,|b| \leq 10^\circ$) is estimated to be $3.3 \times 10^{-4}\,\mathrm{photons\,cm^{-2}\,s^{-1}}$ \citep[using COMPTEL and INTEGRAL data,][]{diehl_2006,bouchet_2015}. The total for the whole Galaxy is estimated to be $1.71\pm 0.06 \times 10^{-3}$ and $2.09\pm 0.08 \times 10^{-3}\,\mathrm{photons\,cm^{-2}\,s^{-1}}$ from COMPTEL and INTEGRAL data, respectively (Pleintinger et al. submitted). 
\begin{figure}
\centering
 \includegraphics[width=0.5\textwidth]{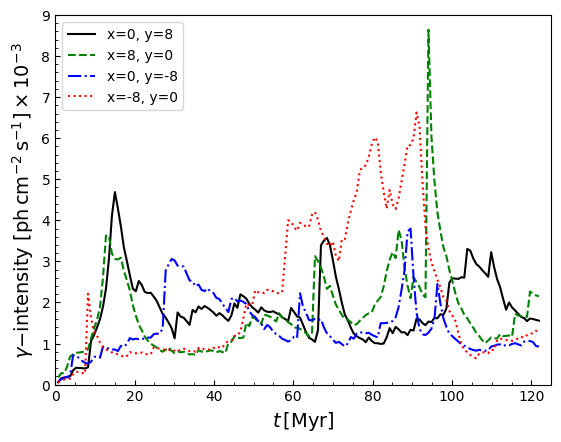}
  \caption{Integrated $\gamma-$ray intensity for the fiducial simulation as a function of time for four different viewing angles.}
\label{fig:integrated_flux_time}
\end{figure}
Fig.\,\ref{fig:integrated_flux_time} shows the integrated flux density from the simulated galaxy for four different viewing angles as a function of time. We calculate the time averaged values of the integrated flux densities from $\sim 69-121$\,Myr and find $1.9, 1.8, 1.3, 3.0\times 10^{-3}\,\mathrm{photons\,cm^{-2}\,s^{-1}}$, respectively. The median integrated flux density values are $1.6, 1.9, 1.1, 2.8\times 10^{-3}\,\mathrm{photons\,cm^{-2}\,s^{-1}}$. The overall minimum and maximum for the integrated flux density values are $6.4 \times 10^{-4}$ and $8.6 \times 10^{-2}\,\mathrm{photons\,cm^{-2}\,s^{-1}}$, respectively. The standard deviation for the integrated flux values, considering all four viewing angles, for $t=69-121$\,Myr is $\sigma=0.2$\,dex which we discuss below in the context of deriving a star formation rate using $^{26}$Al.  

\begin{figure}
\centering
 \includegraphics[width=0.5\textwidth]{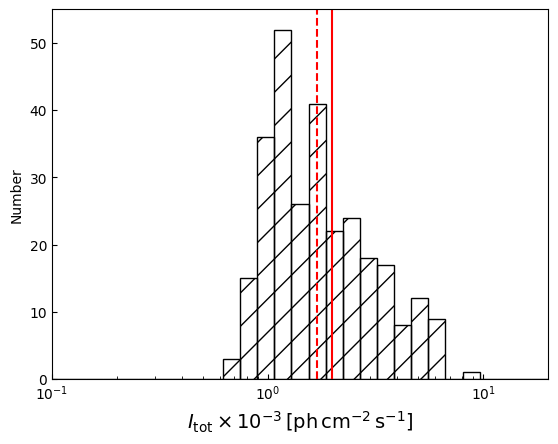}
  \caption{Histogram showing integrated 1809\,keV intensity values for the four different viewing angles between t=69-121\,Myr for the fiducial simulation. The red vertical dashed and solid lines represent the observational values from COMPTEL and INTEGRAL data, respectively.}
\label{fig:histogram}
\end{figure}

Fig.\,\ref{fig:histogram} shows a histogram of the total integrated flux density values for the four viewing angles between $\sim 69-121$\,Myr. This illustrates the variation in the total integrated flux density but that the majority of values are $<2.5 \times 10^{-3}\,\mathrm{photons\,cm^{-2}\,s^{-1}}$.

The total $\gamma-$ray intensity derived from the COMPTEL and INTEGRAL measurements together has been used to infer a star formation rate of $3.8 \pm 2.2 M_\odot \mathrm{yr^{-1}}$ for the Milky Way \citep[][compare with Section\,\ref{sec:form}]{diehl_2006}. Deriving a star formation rate from $^{26}$Al has a number of uncertainties such as the uncertainty of $^{26}$Al yields from massive stars \citep[see][]{voss_2009}, as well as uncertainties relating to the assumed stellar initial mass function \citep[see review by][]{bastian_2010}. Our results, obtained by examining the integrated flux densities and the corresponding emission maps, show that our position in the Galaxy, and specifically our possible close proximity to local sources, could introduce an additional systematic variation of 0.2\,dex on the star formation rate inferred via $^{26}$Al. The star formation rate of $3\,M_\odot \mathrm{yr}^{-1}$ in our simulations was chosen such that the peak of the total 1809 keV line intensity distribution would be similar to the observed values.

Fig.\,\ref{fig:m26} shows the total mass of $^{26}$Al in the galaxy for the fiducial simulation with an average value of $M_{26}=4\,M_\odot$ derived from values between $t=69-121$Myr. \citet{diehl_2017} report a total of $2.0\pm 0.3M_\odot$ for $^{26}$Al using the INTEGRAL/SPI $\gamma$-ray spectrum. The discrepancy is due to the symmetry assumed in the original analysis of the observations. Our simulations should have a more realistic spatial distribution for Milky Way-type galaxies in general. Hence, if the Milky Way currently has a star formation rate of $3\,M_\odot \mathrm{yr}^{-1}$ and displays its most probable 1809 keV intensity, we would correct the steady-state mass of $^{26}$Al for the Milky Way upwards to $\sim 4\,M_\odot$ for such a general galaxy. We discuss alternative scenarios in Section\,\ref{sec:discussion}.

\subsection{Scale height of the galactic disk}
\label{subsec:scale-height}

\subsubsection{Calculation of the scale height}
We calculate the scale height of the $^{26}$Al material both from the 3D grid of density values and from the $\gamma-$ray emission maps for the fiducial simulation and for the higher halo density simulation. The scale height of the material traced by the $^{26}$Al for a generic galaxy may vary from the scale height that we observe from the $\gamma-$ray emission maps which are intrinsically dependent on our position in the Milky Way.

First using the $^{26}$Al densities, we calculate the scale height above, $H^+$, and below, $H^-$, the galactic plane as
\begin{equation}
H^+(R) = \frac{\int \limits_0^\infty \rho_{26}(R)\,z\,dz}{\int \limits_0^\infty \rho_{26}(R)\,dz } \hspace{4mm} \mathrm{and} \hspace{3mm}
H^-(R) =  \frac{\int \limits_{-\infty}^0\rho_{26}(R)\,z\,dz}{\int \limits_{-\infty}^0 \rho_{26}(R)\,dz }
\label{eq:Hplus}
\end{equation}
where the $^{26}$Al densities have been binned as a function of galactocentric radius, $R$, before calculating the scale heights of the disk. Similarly, using the $\gamma-$ray emission maps we can calculate the scale height as a function of galactic longitude as
\begin{equation}
h^+ = \frac{\int \limits_0^{90^\circ} F_\gamma\,b\,db}{\int \limits_0^{90^\circ} F_\gamma\,db } \hspace{5mm} \mathrm{and} \hspace{5mm}
h^- =  \frac{\int \limits_{-90^\circ}^0F_\gamma\,b\,db}{\int \limits_{-90^\circ}^0 F_\gamma\,db }
\label{eq:hplus}
\end{equation}
where $b$ is galactic latitude.

\subsubsection{Comparison of the scale heights}
\begin{figure*}%
	\centering
    \subfigure[ ]{%
        \includegraphics[width=0.5\textwidth]{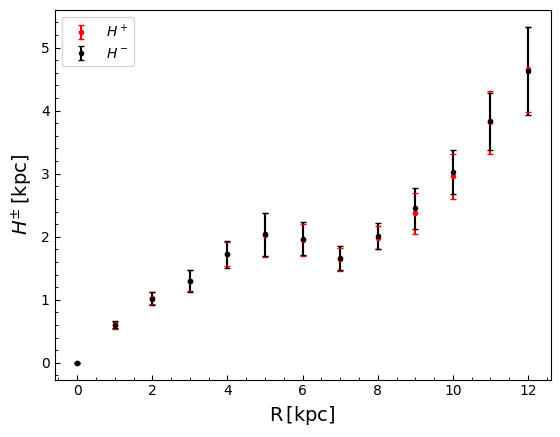}
       	\centering
\label{fig:lh_scale_height}}%
  	    ~
    \subfigure[ ]{%
        \includegraphics[width=0.5\textwidth]{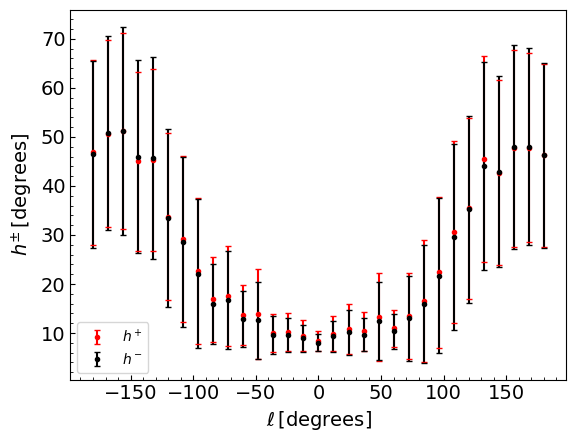}
	\centering
\label{fig:lh_bscale_height}}%
    \caption{(a) plots the scale height of $^{26}$Al in the galaxy for the fiducial simulation as a function of galactocentric radius, $R$. $H^+$ and $H^-$ are the scale heights of the galaxy above and below the disk midplane, calculated using Eq.\,\ref{eq:Hplus}. (b) plots the scale height of the galaxy from the fiducial simulation as a function of galactic longitude, $\ell$. $h^+$ and $h^-$ are the scale heights of the galaxy above and below $0^\circ$ latitude, calculated using Eq.\,\ref{eq:hplus}.} 
    \label{fig:lh}%
\end{figure*}

The scale heights for the fiducial simulation are plotted in Fig.\,\ref{fig:lh}. In Fig.\,\ref{fig:lh_scale_height}  all of the scale heights are averaged between $t=69-121$Myr and the radial bins are 1\,kpc. The scale heights in Fig.\,\ref{fig:lh_bscale_height} have been averaged over the four viewing angles, as well as between $t=69-121$Myr, and the longitude bins are 12$^\circ$. The error bars represent the $1\sigma$ dispersion of the data for the different times and viewing angles.

Fig.\,\ref{fig:lh_scale_height} shows a maximum scale height of $\sim 5$\,kpc at $R=12$\,kpc. There is little difference seen above and below the disk. The scale heights calculated using the emission maps give a maximum scale height of $50^\circ$ at $-150^\circ$ and $150^\circ$ longitude. The scale height of the $^{26}$Al in Fig.\,\ref{fig:lh_bscale_height} is largely symmetric for positive and negative longitude values.

The inner disk, however, has scale heights that are strongly suppressed by the higher halo gas pressure there. This translates directly to the observed latitude extent for longitudes between $-90^{\circ} < \ell < 90^{\circ}$ (compare Fig.\,\ref{fig:lh_scale_height} with Fig.\,\ref{fig:lh_bscale_height}). This is potentially in disagreement with the observations: the maximum entropy map in \citet{pluschke_2001} indicates some level of high latitude emission at all longitudes which we discuss further in Section\,\ref{sec:discussion}.

\subsection{Longitude-velocity diagrams}
\subsubsection{Calculation of line-of-sight velocities}
We are able to examine the kinematics of $^{26}$Al from the simulations. We compute the line-of-sight velocity, $v_\mathrm{26}$, as a function of longitude to compare with the observational results in Fig.\,8 of \citet{kretschmer_2013} which used INTEGRAL data. We calculate the emission weighted line-of-sight velocity of $^{26}$Al with respect to the observer as
\newcommand{\rd}{\mathrm{d}}
\newcommand{\dlp}{\,\mathrm{d} l^\prime}
\newcommand{\dbp}{\,\mathrm{d} b^\prime}
\begin{equation}
v_\mathrm{26}(l,b) = \frac{\int\limits_{b-\Delta b}^{b+\Delta b}\int\limits_{l-\Delta l}^{l+\Delta l} \int\limits_0^\infty \dfrac{\rho_\mathrm{26}}{r^2} v_r \, \rd r \dlp \dbp}{\int\limits_{b-\Delta b}^{b+\Delta b}\int\limits_{l-\Delta l}^{l+\Delta l} \int\limits_0^\infty \dfrac{\rho_\mathrm{26}}{r^2} \, \rd r \dlp \dbp}
\label{eq:lv}
\end{equation}
where $\rho_{26}$ is the mass density of $^{26}$Al, $\ell^\prime$ is galactic longitude and $r$ here represents the distance between the observer and the source of emission. We bin the data in the same way as \citet{kretschmer_2013} in order to make as close a comparison as possible. Therefore, we plot the same longitude range as Fig.\,8 of \citet{kretschmer_2013} with longitude bins of 12$^\circ$ and a latitude range of $\pm 5^\circ$. \citet{kretschmer_2013} found evidence that the gas associated with the 1809\,keV emission is moving faster than would be expected from Galactic rotation alone as shown by the combination of INTEGRAL and CO data plotted in their Fig.\,8. For $|\ell|>10^\circ$ the gas traced by $^{26}$Al is moving $\sim 100-200\, \mathrm{km\,s^{-1}}$ faster than the molecular gas.

The observations of $^{26}$Al observe the emission in the Galaxy up to $10^5$yr ago (considering the light crossing time of photons in the Milky Way). We do not consider this effect for two reasons. First, Fig.\,10 from \citet{kretschmer_2013} indicates that emission on the far side of the Galaxy would contribute $\sim10\%$ due to geometric dilution and so we would only need to consider up to $10^4$\,yr ago (across the Galaxy). Second, superbubbles are not thought to evolve significantly on timescales of $10^4$\,yr which implies that the effect would be small and therefore for simplicity we examine specific instances in time \citep[Figs.\,3-4,][]{krause_2013}.

\subsubsection{Comparison of longitude-velocity diagrams with observational results}

\begin{figure*}
\centering
 \includegraphics[width=\textwidth]{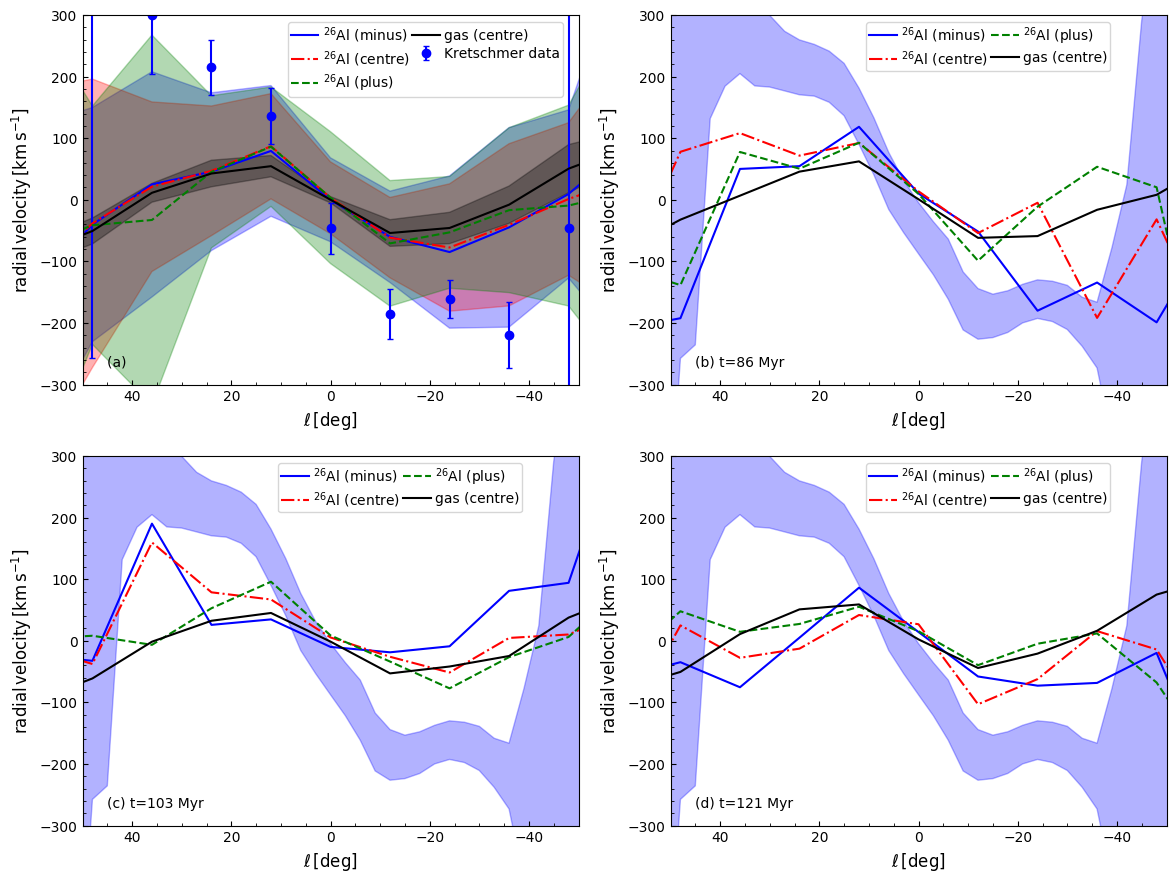}
  \caption{Radial velocities with respect to the observer plotted as a function of galactic longitude, derived from the $^{26}$Al tracer fluid and gas density. (a) shows the radial velocities that have been averaged between $t=69-121\,$Myr. The solid black line plots the time averaged radial velocity for the cold gas in the fiducial simulation. The solid blue, red and green lines plot the time averaged radial velocities for $^{26}$Al and represent the three different simulations where the superbubbles are injected behind, on and ahead of the spiral arms in the galaxy, respectively. The shaded blue, red and green regions correspond to 2$\sigma$ variation of these data. The longitude bin size used is 12$^\circ$.  The blue data points are the $\gamma$-ray observational data from \citet{kretschmer_2013}. (b)-(d) show individual snapshots of the radial velocities for $t=86,103$ and 121\,Myr, respectively. Here the blue shaded region now represents the 1$\sigma$ uncertainty from the observational data.}
\label{fig:lv-panel}
\end{figure*}

We present longitude-velocity diagrams using our 3 simulations in Fig.\,\ref{fig:lv-panel}. For Fig.\,\ref{fig:lv-panel}(a) the data has been averaged between $t=69-121$\,Myr. The line-of-sight velocities as a function of galactic longitude for the cold gas are plotted as the solid black line (which uses $\rho$ rather than $\rho_\mathrm{26}$ in Eq.\,\ref{eq:lv}). The maximum radial velocity of the cold gas shown is $\sim 50\mathrm{km\,s^{-1}}$, which broadly agrees with the CO observations of \citet{dame_2001} shown in Fig.\,8 of \citet{kretschmer_2013}. The grey shaded represents the 2$\sigma$ deviation of the simulation data and indicates that the line-of-sight velocities associated with the cold gas do not vary significantly with time. We have overplotted the $\gamma$-ray observational data from \citet{kretschmer_2013} as blue dots for comparison where the error bars represent the $1\sigma$ statistical uncertainty associated with the observations.

The time-averaged line-of-sight velocities for $^{26}$Al from our simulations are also shown in Fig.\,\ref{fig:lv-panel}(a). The line-of-sight velocities for the three different simulations with the superbubbles injected behind, on and ahead of the spiral arms are plotted as the solid blue, red and green lines, respectively. Thus, the coloured shaded regions in Fig.\,\ref{fig:lv-panel}(a) are the associated 2$\sigma$ deviation of the simulation data. These shaded regions represent the frequency of certain values rather than uncertainties. Evidently, the line-of-sight velocities associated with the $^{26}$Al as observed would vary with time, would we observe $\sim$1-100 Myr later. In contrast, the cold gas shows far less variability. This is understandable since massive star formation and superbubbles evolve on much shorter timescales in comparison to the cold gas.

In Fig.\,\ref{fig:lv-panel}(a) the maximum velocity associated with $^{26}$Al ($\sim100\mathrm{km\,s^{-1}}$) from any of the simulations is compatible with the large observed velocities found by \citet{kretschmer_2013} within the uncertainties (the error bars and shaded blue region in Fig.\,8 of \citet{kretschmer_2013} represent 1$\sigma$ error bars; thus our results agree within 2$\sigma$ from the simulations). Figs.\,\ref{fig:lv-panel}(b)-(d) present longitude-velocity diagrams for a number of specific times. It is evident that certain snapsnots look more similar to the observations than others. For instance, in Fig.\,\ref{fig:lv-panel}(b), at $t=86$\,Myr higher velocities are found than for $t=121$\,Myr (Fig.\,\ref{fig:lv-panel}(d)). The blue shaded region in Figs.\,\ref{fig:lv-panel}(b)-(d) represent the 1$\sigma$ uncertainties for the observational data. 

 As described in Section\,\ref{subsec:sb-inj}, the difference between the three simulations is the injection position of the superbubbles. Despite this difference, the simulations display similar radial velocities. This is in contrast to the explanation put forth by \citet{kretschmer_2013} which we discuss below.

One possible explanation why the three simulations result in similar radial velocities relates to our simulation set-up. The offset from the spiral arms that we consider may not be sufficiently large, despite the fact that our choice of 0.1 radians as the offset was physically motivated (the young cluster begins on the spiral arm and then may drift away from it before forming a superbubble $\sim 5$\,Myr later). The fact that our mean maximum velocities are not quite as high as the $\sim 300\mathrm{km\,s^{-1}}$ from the observations might be related to dynamic effects due to our superbubble implementation. Namely, the random injection positions of our superbubbles are decided based on the initial maximum density of the spiral arm pattern and then rotated using the pattern speed to the corresponding appropriate injection time. As such these positions may not necessarily correspond to regions of maximum density during the simulation due to the dynamic effect of past superbubbles which may have been injected nearby.

Another possible physical explanation is that we do not consider the effect of cosmic rays produced in superbubbles \citep[as in][]{girichidis_2016}. Cosmic rays contribute via pressure which could cause the superbubbles to expand quicker and produce higher velocities more in line with the INTEGRAL results.

The final point to note is that the magnitudes of the radial velocities at $\pm 24^\circ$ in Fig.\,8 of \citet{kretschmer_2013} differ by $\sim 100\mathrm{km\,s^{-1}}$. Similar differences at high and low longitudes are seen at various times in our simulation results, such as at $t=86\,$Myr in Fig.\,\ref{fig:lv-panel}(b). These differences and their time variability can very naturally be explained by the varying positions of the superbubbles throughout the simulated galaxy and the observed Milky Way. We also investigated the effect of using smaller longitude bin sizes and found little difference. 

\section{Discussion}
\label{sec:discussion}
\subsection{Comparison with higher halo density simulations}
In this section we briefly discuss 3 simulations which investigate the effect of increasing the halo density with a decreased scale radius in comparison to the main setup (as described in Section\,\ref{sec:form}). This alternative setup is motivated by the current uncertainty regarding the structure of the Milky Way's hot-gas halo \citep{gupta_2017,bregman_2018}. \citet{bregman_2018} report evidence for a small scale height of only 2.5\,kpc for the Milky Way from X-ray observations. They suggest that the halo density outside of this small core could be consistent with a simple power law. The concentration factor implied by such a small scale height would be significantly different from that of a standard NFW halo. \citet{taylor_2016} have recently determined the concentration factor to be $c_{200}=12$, which yields $r_\mathrm{s} = 21.5$\,kpc. 

We include 3 simulations here (which as for the main setup inject the superbubbles behind, on and ahead of the spiral arms) with $\rho_\mathrm{halo,0} =1.67\times 10^{-28}\mathrm{g\,cm^{-3}}$ used in Eq.\,\ref{eq:rho-halo}, $r_\mathrm{s} = 21.5$\,kpc and $c_{200}=12$.  Such a gas halo would contribute $1.35\times 10^{11}\,M_\odot$ in baryonic mass to the Milky Way's dark matter halo, which is close to the estimated ``missing baryonic mass" \citep[][their Table 5]{gupta_2017}. Increasing the halo density will affect the $^{26}$Al scale heights of the galaxies (the corresponding density plots are given in Appendix\,\ref{appendix:high_rho}). The other diagnostics for the simulations remained very similar irrespective of the setup.

\begin{figure*}%
	\centering
    \subfigure[ ]{%
        \includegraphics[width=0.5\textwidth]{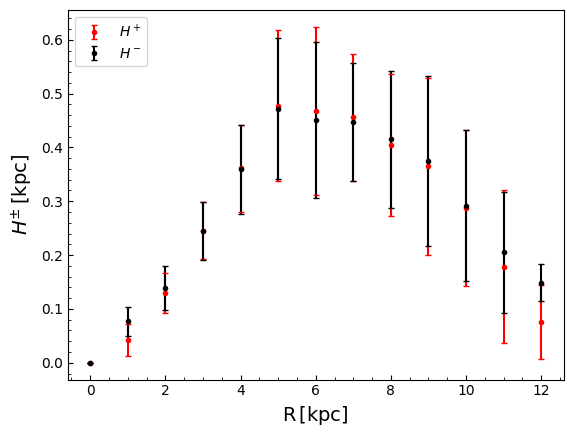}
       	\centering
\label{fig:scale_height}}%
  	    ~
    \subfigure[ ]{%
        \includegraphics[width=0.5\textwidth]{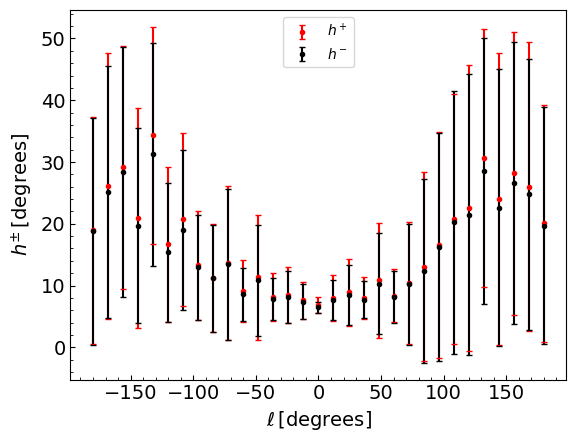}
	\centering
\label{fig:bscale_height}}%
    \caption{ (a) and (b) plot the same scale heights as Fig.\,\ref{fig:lh}(a) and (b) but for the simulation setup with $\rho_\mathrm{halo,0}=1.67\times 10^{-28}\mathrm{g\,cm^{-3}}$.} 
    \label{fig:high_scale_height}%
\end{figure*}

The scale heights for the higher halo density simulation are plotted in Fig.\,\ref{fig:high_scale_height}. As for Fig.\,\ref{fig:lh} the scale heights in Fig.\,\ref{fig:scale_height} are averaged between $t=69-121$\,Myr and the radial bins are 1\,kpc. Similarly, the scale heights in Fig.\,\ref{fig:bscale_height} have been averaged over the four viewing angles, as well as between $t=69-121$Myr, and the longitude bins are 12$^\circ$. The error bars represent the $1\sigma$ dispersion of the data for the different times and viewing angles.

The $^{26}$Al scale height turns out to be a strong function of the halo density, varying by a factor of $\sim$10 between the simulations with low and high halo density. The simulation with high halo density (motivated by the limit allowed by putting the cosmologically expected number of baryons into the gaseous halo) is clearly ruled out by comparison to the maximum entropy map of \citet{pluschke_2001}, which shows significant emission up to latitudes of $\pm50^{\circ}$. With the choice of the halo density constrained by X-ray observations of \citet{gupta_2017}, we find a $^{26}$Al scale height of $50^{\circ}$ at $|\ell|\gtrsim90^{\circ}$ for the lower halo density simulations.

\section{Summary and Conclusions}
\label{sec:conclusions}

In this paper we have investigated the kinematic and morphological imprints that superbubbles have left in the COMPTEL and INTEGRAL observations. We ran 3D hydrodynamic simulations of a spiral galaxy, with physical properties similar to the Milky Way. We investigated the effect of varying the injection position of the superbubbles relative to the spiral arms of the galaxy. Using a tracer fluid to represent the radioisotope $^{26}$Al we produced all sky synthetic $\gamma-$ray maps and, for the first time from 3D simulations, longitude-velocity diagrams. 

\subsection{Derived star formation rate uncertainty}
We find that the $\gamma-$ray emission derived from the simulations is both temporally and spatially very variable, even when the observer's location is constrained to lie on the galactocentric circle, at a distance that corresponds to the solar system's location in the Milky Way. In the simulations as much as 80\% of the 1809\,keV intensity can be local foreground. This supports earlier conjectures that, for the Milky Way, much and especially high-latitude emission can be foreground that varies on a Myr timescale.

This variability affects the measurement of the star formation rate of the Milky Way via $^{26}$Al. Our simulations assume a rate of $3\,M_\odot \mathrm{yr}^{-1}$ tuned such that the most probable 1809\,keV line flux falls within the observed range of values. If this is the situation in the Milky Way, we would recover a steady-state $^{26}$Al mass for the Milky Way of 4\,$M_\odot$. 

We derive an additional uncertainty of 0.2\,dex on the star formation rate, or the total mass of $^{26}$Al, of a Milky Way-type galaxy inferred from $\gamma-$ray emission maps by examining the integrated flux density values from our simulations as a function of time and viewing angle in the galaxy. Taking into account this natural fluctuation, the $^{26}$Al measurement can also be explained by a star formation rate as low as  $1.65 M_\odot \mathrm{yr}^{-1}$ as recently found by \citet{licquia_2015}, who use a Kroupa initial mass function. The steady state $^{26}$Al mass would be correspondingly lower. The relatively high star formation rate derived from the 1809\,keV line intensity would then mean that the Milky Way happens to be in a state with a lot of foreground at the position of the Sun, which is not the most probable, but still a plausible situation.

\subsection{Longitude-velocity diagrams}
\citet{kretschmer_2013} found that the material responsible for the 1809\,keV line emission from INTEGRAL observations has higher line-of-sight velocities than the cold gas traced by CO. They postulated that superbubbles driven by massive stars could be responsible for this difference in velocities as the superbubbles preferentially expand rapidly ahead of the spiral arms \citep{krause_2015} which we investigated in our simulations. 

We confirm that superbubbles, irrespective of their injection relative to the spiral arms, can produce excess Doppler shifts of up to 200 $\mathrm{km\,s^{-1}}$ which is consistent with the INTEGRAL observations, within the uncertainties. On the other hand, we do not find a systematic trend for the Doppler shifts when changing the relative locations of superbubble injections from the leading to the trailing edge of the spiral arm, as suggested would occur by \citet{krause_2015}. The superbubbles in our simulations displace the gas rather symmetrically, so that a systematic effect on $^{26}$Al velocities does not occur. The dense molecular gas which would interact with the superbubbles, as invoked in the model by \citet{krause_2015}, is also however, not present in our simulations. A more self-consistent connection of superbubbles with star formation may additionally be important. 

The large time variability we find in our synthetic longitude-velocity diagrams suggests that not much of a systematic effect may be necessary to explain the observations, as we can only observe one snapshot of a highly variable system. We cannot exclude that the symmetric Doppler shift of the 1809\,keV line observed in the Milky Way is due to a particular configuration of the ISM near the Sun. It is also possible, as noted above, that dense molecular gas or cosmic rays (not included in our simulations) play an important role in contributing to the systematic velocity shifts. We did not include dense molecular gas as this would have required including a chemical network and higher spatial resolution to resolve molecular cloud scales which would have been significantly more computationally expensive.

\subsection{Scale height of $^{26}$Al in the galaxy}
We examined the scale height of the $^{26}$Al in the galaxy as a function of galactocentric radius and also as a function of longitude. We find that the galaxy from the fiducial simulation has a maximum $^{26}$Al scale height of $\sim$5\,kpc. This translates into a maximum scale height of $50^{\circ}$ as seen by an observer located at a galactocentric radius of 8\,kpc, similar to the position of the solar system in the Milky Way. We find that the $^{26}$Al scale height is coupled to the gas density in the Galactic halo. Using observed halo densities produces $^{26}$Al scale heights in broad agreement with the $\gamma$-ray observations, whereas a much higher halo density results in scale heights that are unrealistically small.

While we defer a detailed comparison to the observational data to a dedicated future publication, we find that the scale heights are in broad agreement with the maximum entropy map of \citet{pluschke_2001}, except that we predict a much lower scale height within 90$^{\circ}$ of the galactic centre. This is a direct consequence of the higher halo pressure in our hydrostatic setup. One may speculate, if the density in the inner parts of the Milky Way was reduced by some process, such that the superbubbles encountered less ram pressure, they would thus expand more easily into the halo there. The Fermi bubbles \citep{su_2010} are known to have produced such a low density region in the centre of the Milky Way, but it would hardly be big enough to explain our findings.

Overall, $^{26}$Al has been show to be a sensitive diagnostic for the dynamics of superbubbles and their connection to the gaseous halo of the Milky Way.

\section*{Acknowledgements}
This work has made use of the University of Hertfordshire's high-performance computing facility. DRL acknowledges funding from the Irish Research Council. We thank the anonymous reviewer for their constructive comments.

\appendix

\section{$^{26}$Al tracer}
\label{appendix:isotope}
\begin{figure}
\centering
 \includegraphics[width=0.5\textwidth]{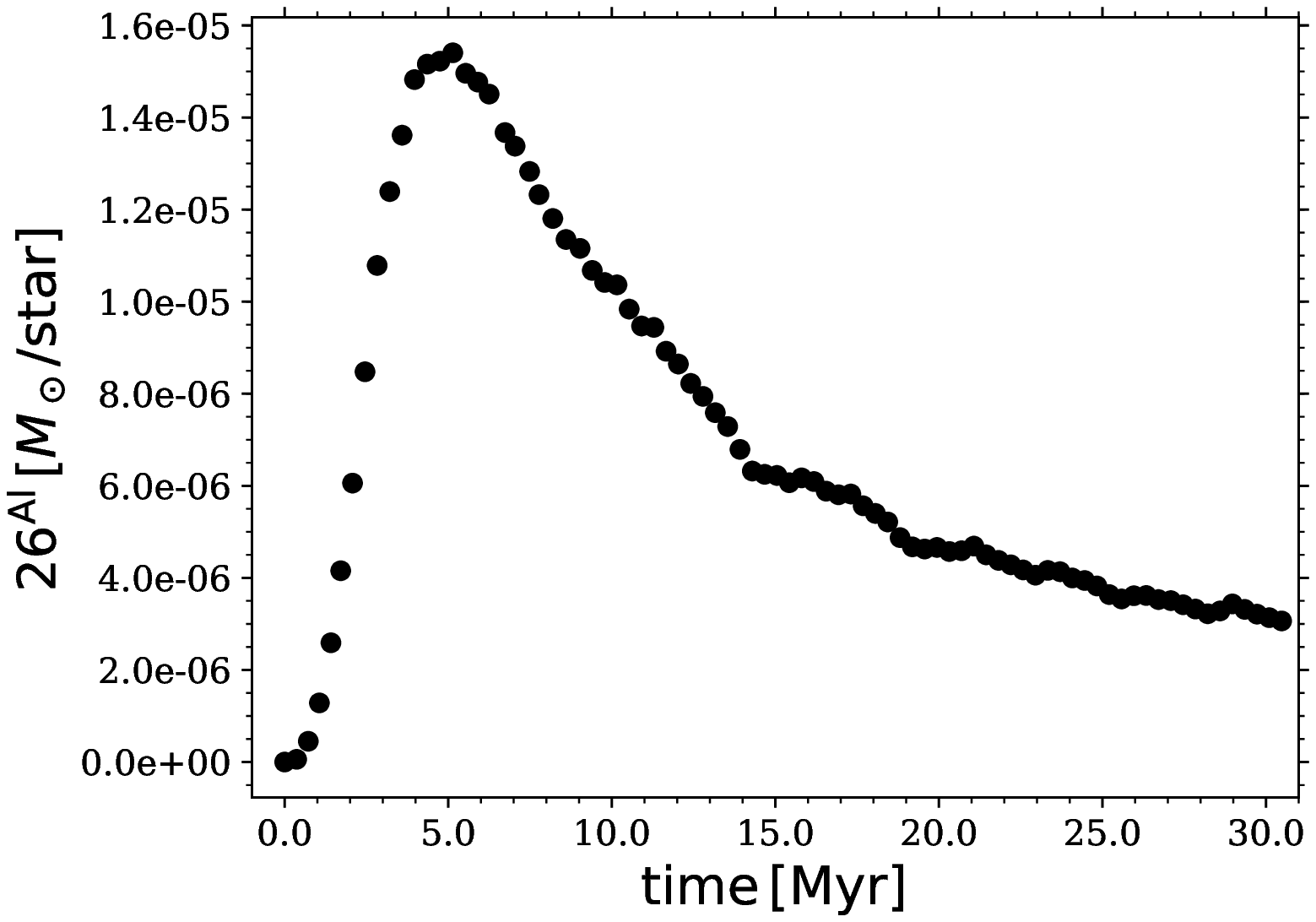}
  \caption{Time profile of the mass injection rate of $^{26}$Al per massive star for our simulations convolved with the radioactive decay of $^{26}$Al.}
\label{fig:isotope}
\end{figure}

The radioisotope $^{26}$Al is used in our simulations as a tracer of massive star formation and to compare with observations, as described in Section\,\ref{subsection:isotope}. Fig.\,\ref{fig:isotope} shows the mass injection rate per massive star as a function time from Eq.\,\ref{eq:m26} (divided by $10^4$), including the effects of radioactive decay from an idealised simulation containing one superbubble. We can compare Fig.\,\ref{fig:isotope} with the solid line in the upper panel of Fig.\,3 from \citet{voss_2009}. We match the increase from $0-5$\,Myr to the maximum value of $\sim1.5\times 10^{-5}M_\odot\,\mathrm{yr}^{-1}$ and the subsequent gradual decline quite well. Note, we do not reproduce the slight bump at $\sim 17$\,Myr. Since we do not trace the first 10\,Myr of the superbubbles as described in Section\,\ref{subsec:sb-inj} we also begin the injection of $^{26}$Al at t=10\,Myr in Eq.\,\ref{eq:m26}.

\section{$\gamma$-ray emission maps}
\label{appendix:maps}
\begin{figure*}
\centering
 \includegraphics[width=\textwidth]{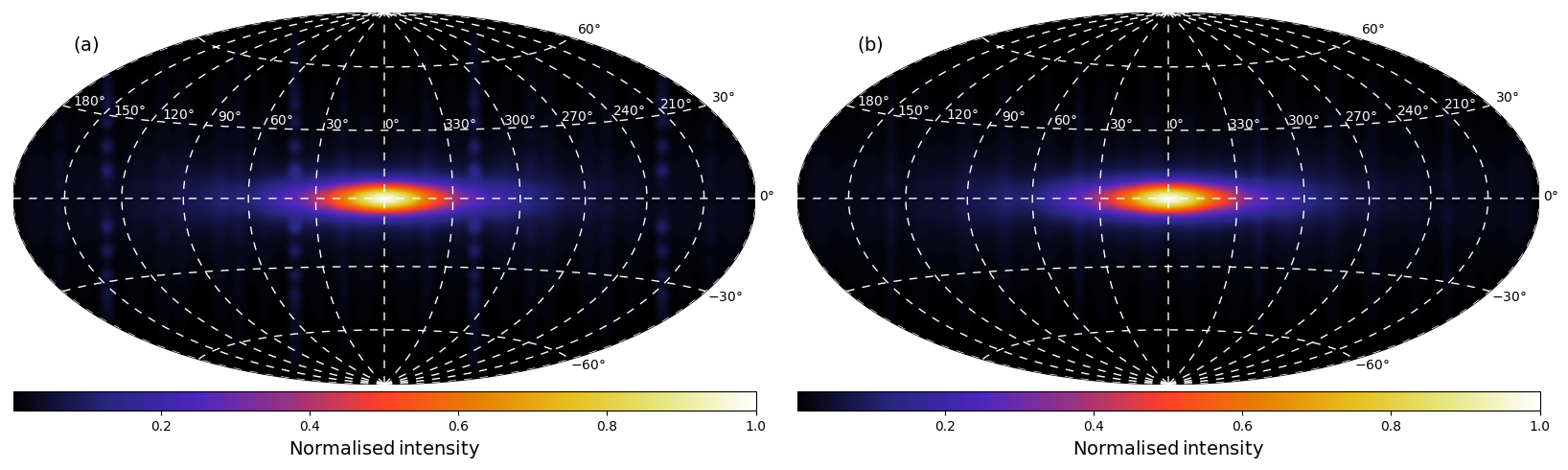}
  \caption{Emission maps showing in (a) the full $\gamma$-ray emission map and in (b) the map with emission within 0.2\,kpc of the observer excluded. Note the emission in this map has been normalised and uses the initial gas density profile of the galaxy. It is intended merely to clearly show the aliasing.}
\label{fig:full_emission}
\end{figure*}

Transforming from the 3D Cartesian grid to a Hammer projection map can create misleading aliasing which is most noticeable when there is significant emission close to the observer. Fig.\,\ref{fig:full_emission}\,(a) shows the full normalised emission using the initial gas mass density profile of the galaxy which varies relatively smoothly (rather than the $^{26}$Al mass density) in comparison to Fig.\,\ref{fig:full_emission}\,(b) which excludes any emission within 0.2\,kpc of the observer. This results in 95\% of the emission being plotted with nearby emission contributing most to this issue. Therefore, when viewing the $\gamma$-ray emission maps in Section\,\ref{subsec:maps}, it is worth bearing in mind that the pattern observed in Fig.\,\ref{fig:full_emission}\,(a) will, to some extent, be superimposed on the maps. Note, this issue is position dependent such that if there is less nearby $\gamma$-ray emission due to the observer's specific location in the Galaxy this effect will be reduced. This dependence is why we decided to show the full emission maps in Section\,\ref{subsec:maps}, as it is difficult to remove these artefacts in any systematic way.

\begin{figure*}
\centering
 \includegraphics[width=\textwidth]{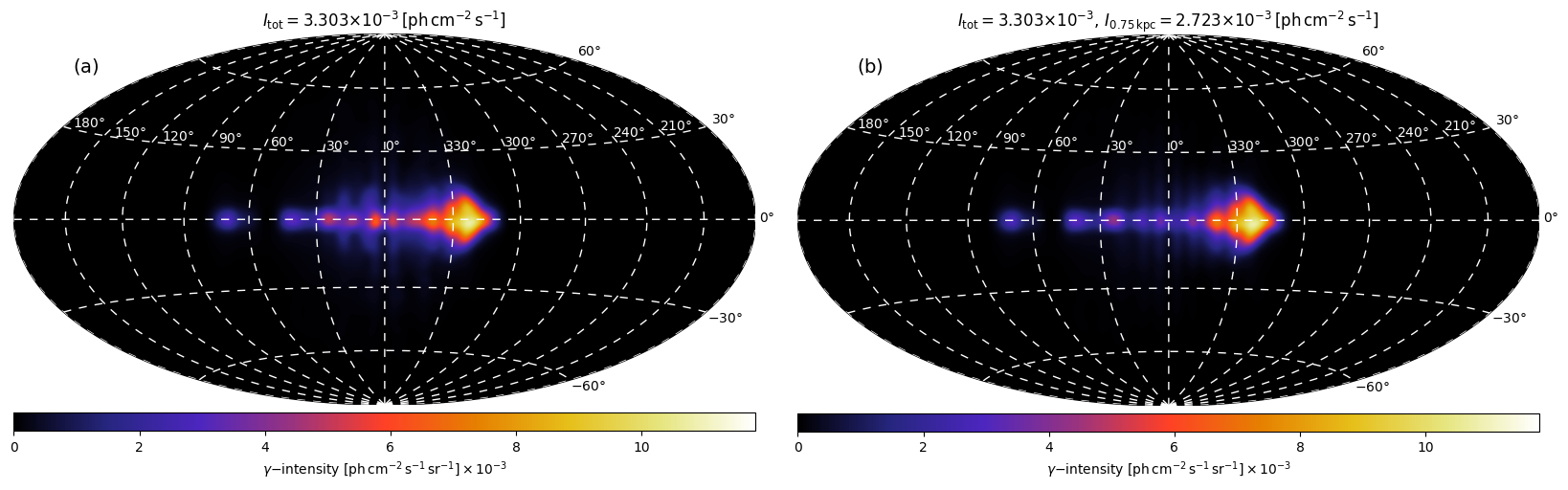}
  \caption{Emission maps showing in (a) the full $\gamma$-ray emission map and in (b) a map with emission within 0.75\,kpc of the observer excluded.}
\label{fig:some_emission}
\end{figure*}

We also include Fig.\,\ref{fig:some_emission} which uses the $^{26}$Al density to derive the $\gamma-$ray emission and shows a comparison of a full emission map versus a map excluding emission within 0.75\,kpc of the observer which includes $\sim 90\%$ of the emission. This data is from the fiducial simulation at $t=103.8$\,Myrs. This plot illustrates for one specific instance in time the effect of excluding nearby emission using the $^{26}$Al flux densities which for this example is moderately small.

\section{Time variability}
\begin{figure*}
\centering
 \includegraphics[width=\textwidth]{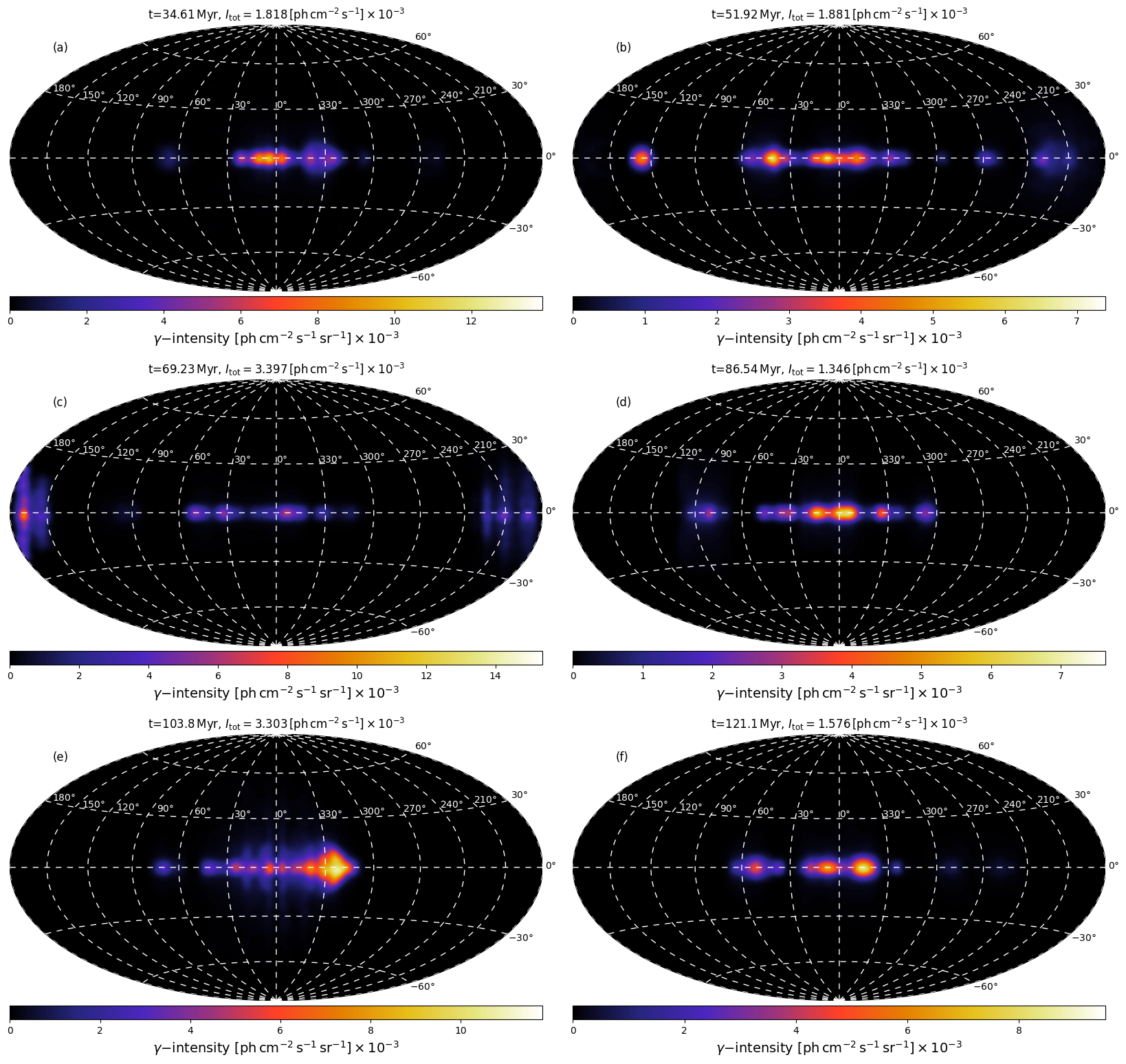}
  \caption{$\gamma-$ray emission maps for the fiducial simulation for different times.}
\label{fig:panel_time}
\end{figure*}

Here we include two plots which show the time variability shown in the emission maps and the longitude-velocity diagrams. Fig.\,\ref{fig:panel_time} plots emission maps from $\sim 34-121\,$Myr with a time interval of $\sim 17\,$Myr. At 34\,Myr the emission is relatively compact as the superbubbles have not had time to significantly perturb the disk. At 51\,Myr individual superbubbles have begun to interact with each other and then by 69\,Myr these have merged to form larger low density regions along the spiral arm of the galaxy. Considering Fig.\,\ref{fig:panel_time} it is immediately evident that the view of the galaxy is time variable on the 17\,Myr timescale considered here.

\section{Higher halo density simulations}
\label{appendix:high_rho}
Many of the quantities that we examined for the simulations gave similar results irrespective of the halo mass density profile used. Here we present the density plots for the higher halo density simulations since the density structures differ significantly to the lower halo density simulations.

\begin{figure*}
\centering
 \includegraphics[width=\textwidth]{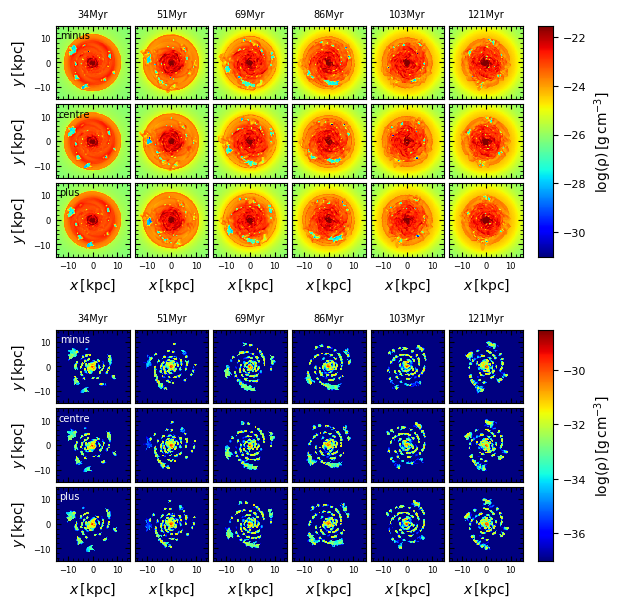}
  \caption{Mass density plots of the 3 simulations with the higher halo density ($\rho_\mathrm{halo,0}=1.67\times 10^{-28}\mathrm{g\,cm^{-3}}$) as a function of time, similar to Fig.\,\ref{fig:rho-panel}.}
\label{fig:apanel-rho}
\end{figure*}

\begin{figure*}
\centering
 \includegraphics[width=\textwidth]{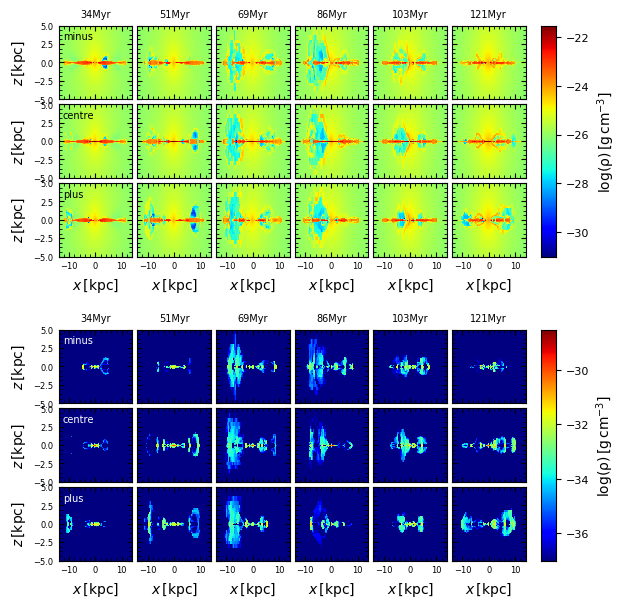}
  \caption{Mass density plots of the 3 simulations with the higher halo density ($\rho_\mathrm{halo,0}=1.67\times 10^{-28}\mathrm{g\,cm^{-3}}$) as a function of time, similar to Fig.\,\ref{fig:rho-panel-slice}. Note, the extent of the $z$-axis is different to Fig.\,\ref{fig:rho-panel-slice}.}
\label{fig:apanel-rho-slice}
\end{figure*}

We have plotted the density profile as a function of time in Figs.\,\ref{fig:apanel-rho}-\ref{fig:apanel-rho-slice} for the higher halo density setup. Changing the initial value of the halo density via $\rho_\mathrm{halo,0}$ changes the long term density profile of the disk as can be seen by comparing Figs.\,\ref{fig:apanel-rho}-\ref{fig:apanel-rho-slice} with Figs.\,\ref{fig:rho-panel}-\ref{fig:rho-panel-slice}. For the lower halo density case (Figs.\,\ref{fig:rho-panel}-\ref{fig:rho-panel-slice}) the superbubbles have expanded more both within the disk and above it. The material above the disk from the superbubbles remains higher in density than the surrounding halo throughout the simulation. In comparison, for the higher halo density case (Fig.\,\ref{fig:apanel-rho}-\ref{fig:apanel-rho-slice}) the influence of the superbubbles is more limited. The expanding bubbles above the disk in Fig.\,\ref{fig:apanel-rho} are pushing into higher density halo material and remain confined to $z\lesssim4\,$kpc. 

Despite these evident differences in the cold gas density structures between the two setups there is little difference in many of the quantities derived from the $^{26}$Al densities, such as the cumulative $\gamma$-ray flux density as a function of distance from the observer, the $\gamma$-ray emission maps and the longitude-velocity diagrams which are not shown here.

\newcommand\aj{AJ} 
\newcommand\actaa{AcA} 
\newcommand\araa{ARA\&A} 
\newcommand\apj{ApJ} 
\newcommand\apjl{ApJ} 
\newcommand\apjs{ApJS} 
\newcommand\aap{A\&A} 
\newcommand\aapr{A\&A~Rev.} 
\newcommand\aaps{A\&AS} 
\newcommand\mnras{MNRAS} 
\newcommand\pasa{PASA} 
\newcommand\pasp{PASP} 
\newcommand\pasj{PASJ} 
\newcommand\solphys{Sol.~Phys.} 
\newcommand\nat{Nature} 
\newcommand\bain{Bulletin of the Astronomical Institutes of the Netherlands}
\newcommand\memsai{Mem. Societa Astronomica Italiana}
\newcommand\apss{Ap\&SS} 
\newcommand\qjras{QJRAS} 
\newcommand\pof{Physics of Fluids}
\newcommand\grl{Geophysical Research Letters}
\newcommand\physrep{Physics Reports}
\newcommand\gca{Geochimica et Cosmochimica Acta}
\bibliographystyle{mn2e}
\bibliography{../../donnabib}

\label{lastpage}

\end{document}